\documentclass[english,aps,prl,amsmath,amssymb,twocolumn,letterpaper,citeautoscript,superscriptaddress,longbibliography]{revtex4-1}

\usepackage{mathtools} 

\usepackage[normalem]{ulem} 
\usepackage{svg}
\usepackage{comment}
\usepackage{amsmath}
\usepackage{physics}
\usepackage{graphicx}
\usepackage{babel}
\usepackage{amstext}
\usepackage{esint}
\usepackage{cases}
\usepackage[bookmarks=false,linkcolor=blue,urlcolor=blue,colorlinks,citecolor=blue]{hyperref}
\usepackage{pdfpages}	
\usepackage{pgffor}

\makeatletter
\AtBeginDocument{\let\LS@rot\@undefined} 
\makeatother

\begin{document}

\title{Microwave spectroscopy of Majorana vortex modes} 

\author{Zhibo Ren}

\affiliation{Department of Physics and Astronomy, Purdue University, West Lafayette, Indiana 47907 USA}

\author{Justin Copenhaver}

\affiliation{Department of Physics and Astronomy, Purdue University, West Lafayette, Indiana 47907 USA}
\affiliation{Department of Physics, University of Colorado, Boulder, CO 80309, USA}

\author{Leonid Rokhinson}

\affiliation{Department of Physics and Astronomy, Purdue University, West Lafayette, Indiana 47907 USA}

\author{Jukka I. V\"ayrynen}

\affiliation{Department of Physics and Astronomy, Purdue University, West Lafayette, Indiana 47907 USA}

\date{\today}
\begin{abstract}
The observation of zero-bias conductance peaks in vortex cores of certain Fe-based superconductors has sparked renewed interest in vortex-bound Majorana states. 
These materials are believed to be intrinsically topological in their bulk phase, thus avoiding potentially problematic interface physics encountered in superconductor-semiconductor heterostructures. 
However, progress toward a vortex-based topological qubit is hindered by our inability to measure the topological quantum state of a non-local vortex Majorana state, i.e., the charge of a vortex pair. 
In this paper, we theoretically propose a microwave-based charge parity readout of the Majorana vortex pair charge. 
A microwave resonator above the vortices can couple to the charge allowing for a dispersive readout of the Majorana parity. 
Our technique may also be used in vortices in conventional superconductors and allows one to probe the lifetime of vortex-bound quasiparticles, which is currently beyond existing scanning tunneling microscopy capabilities. 
\end{abstract}
\maketitle

\textbf{\emph{Introduction.~}}
Majorana zero modes were originally proposed within the context of vortices in a topological superconductor (SC)~\cite{Volovik1999,2000PhRvB..6110267R,2001PhRvL..86..268I,2008PhRvL.100i6407F,2010PhRvL.104d0502S}, and have since emerged as a captivating subject of study in the field of superconductivity. 
The recent discovery of zero-bias conductance peaks~\cite{chen2018discrete,doi:10.1126/science.aao1797,PhysRevX.8.041056,Chen_2019,Machida2019,2019NatPh..15.1181K,doi:10.1126/science.aax0274,Liu2020,Kong2021} in the vortex cores of certain Fe-based superconductors~\cite{SHI2017503,doi:10.1126/science.aan4596,Zhang2019,PhysRevB.100.155134,Li_2022}  has sparked renewed interest in vortex Majorana zero modes (MZMs), which are predicted to be bound in these vortices~\cite {2010PhRvL.104d0502S,Qin_2019,10.1093/ptep/ptad084}. 
The inherent topological nature of vortices as excitations within the superconducting condensate gives hope that the bound states hosted by them would be less susceptible to the disorder, unlike Majorana approaches that require engineered interfaces~\cite{PhysRevB.107.245304,PhysRevB.107.245423}. 
The key motivation behind studying MZMs is their predicted non-Abelian braiding statistics and possible use in a topologically protected quantum computer~\cite{2001PhRvL..86..268I,2003AnPhy.303....2K,RevModPhys.80.1083,Grosfeld_2011}.

However, the measurement of the topological quantum state of a non-local vortex MZM remains a challenge, hindering progress toward a vortex-based topological qubit. While it is in principle possible to move the vortices and associated MZMs~\cite{PhysRevLett.98.010506,PhysRevB.101.024514,2021JPhD...54P4003M,PhysRevB.104.104501}, it will be challenging to do this adiabatically for a large vortex, but at the same time fast enough to avoid quasiparticle poisoning, the timescale of which in vortices is currently unknown. 
Alternatively, measurement-based braiding techniques could potentially circumvent the need for moving the MZMs~\cite{PhysRevLett.101.010501}. Non-local conductance~\cite{PhysRevApplied.12.054035,Sbierski_2022} and interferometric~\cite{PhysRevLett.102.216403,PhysRevLett.102.216404} measurements have been suggested as a means to identify and control Majorana vortex modes. Nevertheless, it is important to note that a microwave-based technique would be optimal for achieving fast readout~\cite{PRXQuantum.1.020313,Razmadze2018}.

In this paper, we propose a solution to the measurement problem using microwave (MW) techniques, which have been established and demonstrated as an extremely versatile tool to address electronic systems in various experiments~\cite{2013Natur.499..312B,Woerkom2016,2019PhRvX...9a1010T,hays2020continuous, hays2021coherent, fatemi2021microwave,PhysRevLett.128.197702,arxiv.2208.11198}.  Specifically, we present a microwave-based method for MZM charge parity readout analogous to what has been proposed in different platforms~\cite{Yavilberg2015,2014NatCo...5E4772G,PhysRevB.92.245432,Vayrynen2015,Yavilberg2015,PRXQuantum.1.020313}.

Our approach focuses on studying the coupling between electrons in the Fe-based superconductor and the microwave photons from a resonator positioned above it. By analyzing the frequency-dependent transmission of the resonator, we can achieve a dispersive readout of the non-local vortex Majorana state. We provide the necessary requirements for the resonator quality factor $Q$ to enable the parity readout. Importantly, our technique can also be applied to vortices in conventional superconductors, offering insights into the lifetime and coherent manipulation of vortex-bound quasiparticles, surpassing the capabilities of existing scanning tunneling microscopy.  

\begin{figure}
\centering
\includegraphics[width=1.0\columnwidth]{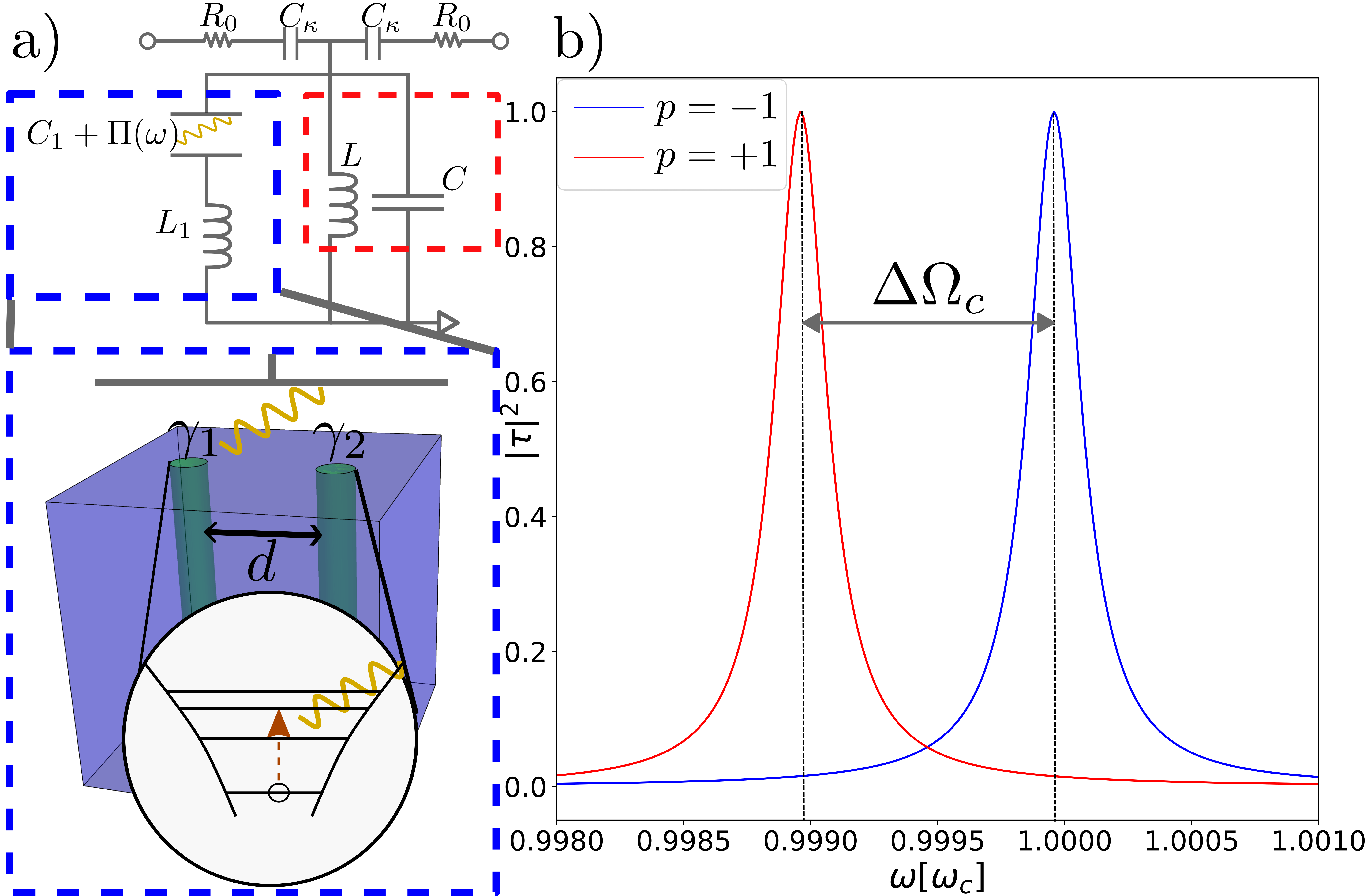}

\caption{ 
Dispersive readout of vortex MZM parity. (a) Schematic circuit model. The red square shows the cavity resonator and the blue squares show the capacitive coupling to the SC vortex state.
The microwave response of the vortex pair, represented by the charge-charge correlation function $\Pi(\omega)$ depends on the MZM parity as described in Eq.~(\ref{eq:correlation function}). 
 (b) transmission vs. frequency in the detuning regime $|E_1-\omega_c| \gg \omega_c\zeta, E_{\text{M}}$. Here $E_1,E_{\text{M}}$ are the energies of the first bound state and Majorana state, $\zeta$ is the dimensionless charge [see above Eq.~(\ref{eq:Qc1})], $\omega_c$ is the resonant frequency of the cavity. The parity readout is measuring $\langle i\gamma_1 \gamma_2 \rangle$, the occupation number parity of the Majorana state on the top surface. 
 We take here the first bound state energy $E_1 \approx 2 \omega_c$, $E_{\text{M}}=0$, $\zeta=0.015$, and
  $\delta \zeta=0.02$. 
 These parameters correspond to critical cavity Q-factor $Q_c\approx10^3$, and in the plot, we take $Q=10^4\gg Q_c$, so these peaks can clearly be resolved.
} \label{fig:system}
\end{figure}

\textbf{\emph{General theory of MW coupling to vortex state.~}}
The interaction between the external electromagnetic field with the charge density of the superconductor results in a MW 
coupling Hamiltonian 
\begin{equation}
    \delta H \cos{\omega t}=  \int d^{3}\mathbf{r} \rho_e(\mathbf{r})  V(\mathbf{r})\cos{\omega t},
    \label{eq:EM coupling}
\end{equation}
where $\rho_e(\mathbf{r})$ is the charge density operator  and  $V(\mathbf{r})\cos{\omega t}$ is the scalar potential of the external electromagnetic field. This electromagnetic field is created by a resonator, which is within a wavelength of the SC surface. Thus the field can be treated in the quasistatic approximation. The sketch of the measurement setup is shown in  Fig.~\ref{fig:system}(a). 

In the static field approximation, screening in the superconductor results in a decay of the field, characterized by a screening length $\lambda_{\text{TF}}$. The scalar potential of the external electromagnetic field can be written as $V(\mathbf{r})=V_0 e^{-\frac{z}{\lambda_{\text{TF}}}}$, where $z$ is the distance from the top surface of the superconductor and $V_0$ is the amplitude of the external field. 

In Eq.~(\ref{eq:EM coupling}), the charge density $\rho_e$ can be expressed as $\rho_e =-\frac{1}{2}e\Psi^\dagger \tau_z \Psi $, where $\Psi  =((c_{\uparrow},c_{\downarrow}),(c_{\uparrow}^\dagger,c_{\downarrow}^\dagger))^T$ is the Nambu field operator  and $\tau_z = \text{diag}(1,1,-1,-1)$. In order to expand the field operator in the exact eigenbasis of the unperturbed Hamiltonian $H_0$, which is given by Eq.~(\ref{eq:3D SC}) below, we define $\Phi_n$ as the spinor wave function of the eigenstate with energy $E_n$, and $\Gamma_n$ as the second quantized annihilation operators of these quasiparticles.

The eigenstates of the system exhibit a particle-hole symmetry (PHS) that is represented by an antiunitary operator $\mathcal{P}$. For each eigenstate $\Phi_n$ with energy $E_n$, there exists another eigenstate $\Phi_{-n} = \mathcal{P}\Phi_n$ with energy $-E_{n}$. The corresponding annihilation operator satisfies $\Gamma_{-n} = \Gamma_n^\dagger$. We consider energies below the SC gap and include the excited vortex-bound states (Caroli-de Gennes-Matricon states). The lowest energy state in the system is the Majorana state $\Phi_\text{M}$ with energy $E_{\text{M}}$, and its corresponding operator is given by $\Gamma_\text{M}=\frac{1}{2}(\gamma_1-i\gamma_2)$, where $\gamma_1$ and $\gamma_2$ are two Majorana operators as shown in Fig.~\ref{fig:system}. We aim to read out the occupation number $n_{\text{M}} = \Gamma_\text{M}^{\dagger} \Gamma_\text{M}  $ [or its parity, $(-1)^{n_\text{M}}= i \gamma_1 \gamma_2$] of this Majorana zero mode.

Expanding the Nambu spinor $\Psi$ in terms of $\Gamma_n$, 
\begin{align}
    \Psi =& \sum_{E_n > 0}  (\Phi_n \Gamma_n+\Phi_{-n} \Gamma_n^\dagger),
    \label{eq:field operator}
\end{align}
the MW coupling~(\ref{eq:EM coupling}) can be written as
\begin{align}
    \delta H & =V_0 \sum_{E_n > 0}q_{n,n}(\Gamma_n^\dagger \Gamma_n-\frac{1}{2}) \label{eq:perturbation} \\
    &
    +\frac{1}{2}V_0 \sum_{E_n > 0}\sum_{m\ne n\atop E_m > 0}[q_{n,m}\Gamma_n^\dagger \Gamma_m \nonumber+q_{n,-m}\Gamma_n^\dagger \Gamma_m^\dagger +H.c.].
\end{align}
Here we introduced the matrix elements of the (surface) charge operator $\hat{q}=2\int d^{3}\mathbf{r} \rho_e e^{-z/\lambda_{\text{TF}}}$, e.g., 
 \begin{equation}
    q_{n,m} =-e\int d^{3}\mathbf{r}  (\Phi_n^\ast \tau_z \Phi_m) e^{-z/ \lambda_{\text{TF}}} .
    \label{eq:matrix element}
\end{equation} 
Because the charge operator preserves PHS, the matrix elements obey the same symmetry, encoded by the relations $q_{n,-m}=-q_{-n,m}^*$ and $ q_{n.-n}=0$.

\textbf{\textit{Microwave readout of Majorana parity.~}}
In circuit quantum electrodynamics~\cite{Blais2020}, the MW coupling between a resonator and the superconductor allows us to read out the Majorana parity~\cite{PhysRevB.92.245432}. The electromagnetic fields induced by the resonator interact with the superconductor in the manner described by the  Hamiltonian $\delta H$,  Eq.~(\ref{eq:perturbation}). This interaction influences the complex transmission coefficient $\tau^{(p)}(\omega)$ that relates the output and input photonic fields of the resonator. 
Under the limit $L_1 C_1\ll LC$ (see Fig.~\ref{fig:system}a) and frequency close to the cavity resonance $\omega_c=1/\sqrt{L C_{\text{tot}}}$, we find 
\begin{equation}
    \tau^{(p)}(\omega)\approx\frac{\kappa}{i(\omega-\omega_c)+\kappa+\frac{ i\omega_c \Pi^{(p)}(\omega)}{2 C_{\text{tot}}}},
    \label{eq:transmission}
\end{equation}
where $\kappa =2 / (C_{\text{tot}} R^*)$ is the escape rate of the cavity, and $p = (-1)^{n_\text{M}}$ denotes the Majorana parity. 
We denote $\Pi^{(p)}(\omega)$  the parity-dependent charge-charge correlation function. In the time domain, it is given by  $\Pi^{(p)}(t)=-\frac{i}{\hbar}\Theta(t)\langle[\hat{q}(t),\hat{q}(0)] \rangle_{p}$, where $\Theta(t)$ is the Heaviside step function. As shown in Fig.~\ref{fig:system}a, $C_{\text{tot}}=C+C_{\text{1}}$, where $C$ and $C_{\text{1}}$ are the capacitance of the resonator and the superconductor, respectively. The resonator is coupled with capacitance $C_{\kappa}$ to the input-output transmission line with resistance $R_0$, and the effective resistance $R^*=\frac{1+\omega_c^2 C^2_\kappa R_0^2}{\omega_c^2 C^2_\kappa R_0}$ 
 incorporates the coupling strength $C_{\kappa}$~\cite{10.1063/1.3010859}. 

The interaction between the resonator and the superconductor induces transitions between the Majorana state and the vortex-bound states localized near the top surface. 
The correlation function $\Pi^{(p)}$ contains information of these transitions and can be written as a sum (from hereon we set $\hbar=1$)
\begin{equation}
\begin{aligned}
&\Pi^{(p)}(\omega)=\sum_{l\ne \pm \text{M}, E_l > 0}\left(\frac{1}{\omega_{l}^{(p)}+\omega+i\delta}+\frac{1}{\omega_{l}^{(p)}-\omega-i\delta}\right) \\
&[|q_{l,+\text{M}}|^2 (n_\text{M}-n_l)-|q_{l,-\text{M}}|^2(n_\text{M}-1+n_l)],
\label{eq:correlation function}
\end{aligned}
\end{equation} 
here $\omega_{l}^{(p)}=E_l+p E_{\text{M}}$ is the transition frequency and  $E_{\text{M}}$, $E_l$,  $n_\text{M}$ and $n_l$  are the energies and occupation numbers of the  Majorana state and the bound state $l$. The infinitesimal level width $\delta > 0$  accounts for causality and $q_{l,\pm \text{M}}$ are the charge matrix elements between bound state $l$ and occupied/unoccupied ($+/- \text{M}$) Majorana state. 
At low temperatures, in the absence of occupied bound states ($n_l=0$), we obtain that $\Pi^{(+)}(\omega)\propto |q_{l,-\text{M}}|^2$ for $n_\text{M}=0$ and $\Pi^{(-)}(\omega)\propto |q_{l,+\text{M}}|^2$ for $n_\text{M}=1$. The unequal charge matrix elements $q_{l,\text{M}},q_{l,-\text{M}}$ and transition frequencies $\omega_{l}^{(p)}$ result in different $\Pi^{(\pm)}(\omega)$, which suggests that the MW coupling can be used to probe the Majorana occupation number  $n_\text{M}$. 

\textbf{\emph{The critical cavity Q-factor.~}}
The parity-dependent correlation function $\Pi^{(p)}$ allows for the microwave readout of MZM  parity based on the transmission [Eq.~(\ref{eq:transmission})]. For simplicity, let us only consider the first vortex-bound state $l=1$ on the surface. Our primary interest lies in the strong coupling regime where the coupling strength $|q_{1,\pm \mathrm{M}}|$ greatly exceeds the level width $\delta$. In this regime,  the transmission $|\tau^{(p)}|^2$ versus $\omega$ displays two parity-dependent peaks at $\omega>0$. Parity readout is contingent upon the ability to distinguish peaks between different parity, which sets limitations for the cavity Q-factor $Q=\frac{\omega_{c}}{\kappa}$. Here, we define a minimum critical cavity Q-factor $Q_c$ required for parity distinction,
\begin{equation}
  Q_c^{-1}=\frac{\Delta\Omega_{c}}{\Omega_c} ,
    \label{eq:def Qc}
\end{equation}
where $\Delta \Omega_{c}=\Omega_{c}^{(+)}-\Omega_{c}^{(-)}$ is the peak seperation of two parities in Fig.~\ref{fig:system}(b), $\Omega_{c}=\frac{1}{2}(\Omega_{c}^{(+)}+\Omega_{c}^{(-)})$ is the average of peak positions, and $\Omega_{c}^{(\pm)}$ are the shifted resonator frequencies of $p=\pm1$ parity~\cite{SuppMat}, i.e., $|\tau^{(p)}(\Omega_{c}^{(p)})|^2=1$. 

The Eq.~(\ref{eq:def Qc}) determines the approximate requirement $Q>Q_c$ for distinguishing the different parity resonances. There are two variables that affect the critical cavity Q-factor $Q_c$: the parity-dependent charge matrix elements $q_{1,\pm\text{M}}$ and the parity-dependent transition energies $\omega_1^{(p)}$ associated with the Majorana energy splitting $E_{\text{M}}$. We define the dimensionless variable $\zeta_{\pm \text{M}}=\sqrt{\frac{U_{\pm \text{M}}}{\omega_c}}$ and the capacitive energy $U_{\pm \text{M}}=\frac{q_{1,\pm \text{M}}^2}{2 C_{\text{tot}}}$. 

In the resonant regime, when the resonator frequency is close to the energy of the first bound state, characterized by $|\omega_c-E_1|\ll \omega_c \zeta$, where $\zeta=\frac{1}{2}(\zeta_{+\text{M}}+\zeta_{-\text{M}})$, the transmission curve of each parity exhibits two peaks of width $\kappa$, separated by $2\omega_c \zeta$. The parity difference causes a shift in the position of the peaks by $\omega_c \delta \zeta-\frac{1}{2} \delta E$, where $\delta \zeta=\zeta_{+\text{M}}-\zeta_{-\text{M}}$ is the dimensionless transition matrix element difference, and $\delta E$ is the change in resonance frequency given by $\delta E=-2 E_{\text{M}}$. It is important to note that these two contributions have opposite effects, which can affect the behavior of the transmission curve in this regime. By setting the shift in peak position equal to the escape rate $\kappa$, we obtain
\begin{equation}
   Q_{c
   }\approx |\frac{\omega_c}{\omega_c \delta \zeta+E_{\text{M}}}|, \qquad |\omega_c-E_1|\ll \omega_c \zeta.
    \label{eq:Qc1}
\end{equation}
It is worth mentioning that $Q_c$ diverges at $\delta \zeta = -E_{\text{M}}/\omega_c$ since the peak position does not shift, and thus parity detection becomes difficult. 

In the detuning regime, where the resonator frequency is significantly detuned from the first bound state's energy ($|E_1-\omega_c| \gg \omega_c\zeta, E_{\text{M}}$), the full expression for $Q_c$ is more complex compared to the resonant regime (for detailed derivation, see Ref.~\cite{SuppMat}). Nevertheless, $Q_c$ can be approximated as:
\begin{subnumcases}{Q_{c} \approx}
   |\frac{E_1^2-\omega_{c}^2}{4\omega_{c} E_1\zeta \delta \zeta}|,&$\frac{\delta \zeta}{ \zeta} \gg \frac{ E_{\text{M}}( E_1^2+\omega_{c}^2)}{(E_1^2-\omega_{c}^2)E_1}$, \label{eq:Qc2a}
   \\
    \frac{(E_1^2-\omega_{c}^2)^2}{4\omega_{c}(E_1^2+\omega_{c}^2)E_{\text{M}}\zeta^2},&$\frac{\delta \zeta}{ \zeta} \ll \frac{E_{\text{M}}( E_1^2+\omega_{c}^2)}{(E_1^2-\omega_{c}^2)E_1}$,\label{eq:Qc2b}
   \label{eq:Qc2}
\end{subnumcases}
where $|E_1-\omega_c| \gg \omega_c\zeta, E_{\text{M}}$. The two different forms highlight the parity readout based on the parity-dependence of the charge matrix element ($\delta \zeta$) or the transition energy ($E_{\text{M}}$). The first form [Eq.~(\ref{eq:Qc2a})] depends only on the change in the dimensionless charge matrix element difference $\delta \zeta$ as the parity-dependent factor, while the second form [Eq.~(\ref{eq:Qc2b})] depends only on the change in transition energy $2E_{\text{M}}$ as the parity-dependent factor. 

\textbf{\emph{Model for Fe-based superconductor.~}}
In order to estimate the feasibility of the parity readout discussed above, we will use a microscopic Hamiltonian to evaluate the transition matrix element between the Majorana state and the vortex-bound states. 

We will analyze a two-band effective BdG model for an Fe-based superconductor \cite{Qin_2019,PhysRevLett.122.187001,PhysRevB.101.020504,doi:10.1126/sciadv.aay0443,HouKlinovaja2021,PhysRevB.105.035128}. The Hamiltonian in the Nambu basis~$\Psi(\textbf{k})=(c_{1\uparrow},c_{1\downarrow},c_{2\uparrow},c_{2\downarrow},c_{1\uparrow}^\dagger,c_{1\downarrow}^\dagger,c_{2\uparrow}^\dagger,c_{2\downarrow}^\dagger)^T$ can be represented as $H_{\text{SC}}=\frac{1}{2} \int dk \Psi^\dagger \mathcal{H}_{\text{SC}} \Psi$, where the BdG Hamiltonian $ \mathcal{H}_{\text{SC}}$ is given by
\begin{equation}
    \mathcal{H}_{\text{SC}}=\begin{pmatrix} H_0(\textbf{k})-\mu&i\Delta_0 \sigma_y\\-i\Delta_0^* \sigma_y&\mu-H_0^*(-\textbf{k}) \end{pmatrix},
    \label{eq:3D SC}
\end{equation}
here $\mu=5$ meV represents the chemical potential, and $\Delta_0=1.8$ meV is the bulk pairing gap. In our lattice model, $H_0(\textbf{k})=\nu \eta_x(\sigma_x\sin k_x a+\sigma_y\sin k_y a+\sigma_z\sin k_z a)+m(\textbf{k})\eta_z$ with $m(\textbf{k})=m_0-m_1(\cos k_x a+\cos k_y a)-m_2 \cos k_z a$, where $\eta_i$ and $\sigma_i$ represent the Pauli matrices that account for the orbital and spin degrees of freedom, respectively \cite{PhysRevLett.122.187001}. In this basis, $\mathcal{P}=\tau_x K$, where $\tau_x$ represents the Pauli matrix that accounts for the particle-hole degrees of freedom and $K$ denotes complex conjugation. In our numerical simulation, we set $\nu=10$ meV and $a =5$ nm (the lattice constant), $m_0=-4 \nu,m_1=-2 \nu$, and $m_2=\nu$ so that the system is in the topological phase \cite{PhysRevLett.122.187001,HouKlinovaja2021} which can have vortex Majorana zero modes.

In the context of our model, we consider the s-wave superconducting pairing potential in the presence of vortices that extend along the z-axis. For a single vortex centered at the origin, the pairing term can be expressed as~\cite{PhysRevLett.77.566}
\begin{equation}
    \Delta_{1-\text{v}}(w)=\Delta_0 \frac{w}{\sqrt{|w|^2+\xi^2}},
    \label{eq: single vortex}
\end{equation}
where $\xi=5$ nm represents the characteristic radius of the vortex and $w=x+iy$. 

In our specific model (shown in Fig.~\ref{fig:system}a), we consider the presence of two vortices, each hosting a pair of MBSs within the Fe-based superconductor. Assuming the vortices are far apart, we can approximate the pairing term as follows,
\begin{equation}
      \Delta_{2-\text{v}}(w)=\Delta_0 \frac{w-w_1}{\sqrt{|w-w_1|^2+\xi^2}}\frac{w-w_2}{\sqrt{|w-w_2|^2+\xi^2}},
      \label{eq: two vortex}
\end{equation}
where $w_1$ and $w_2$ correspond to the respective locations of the two vortices. 

\begin{figure}
\centering
\includegraphics[width=1\columnwidth]{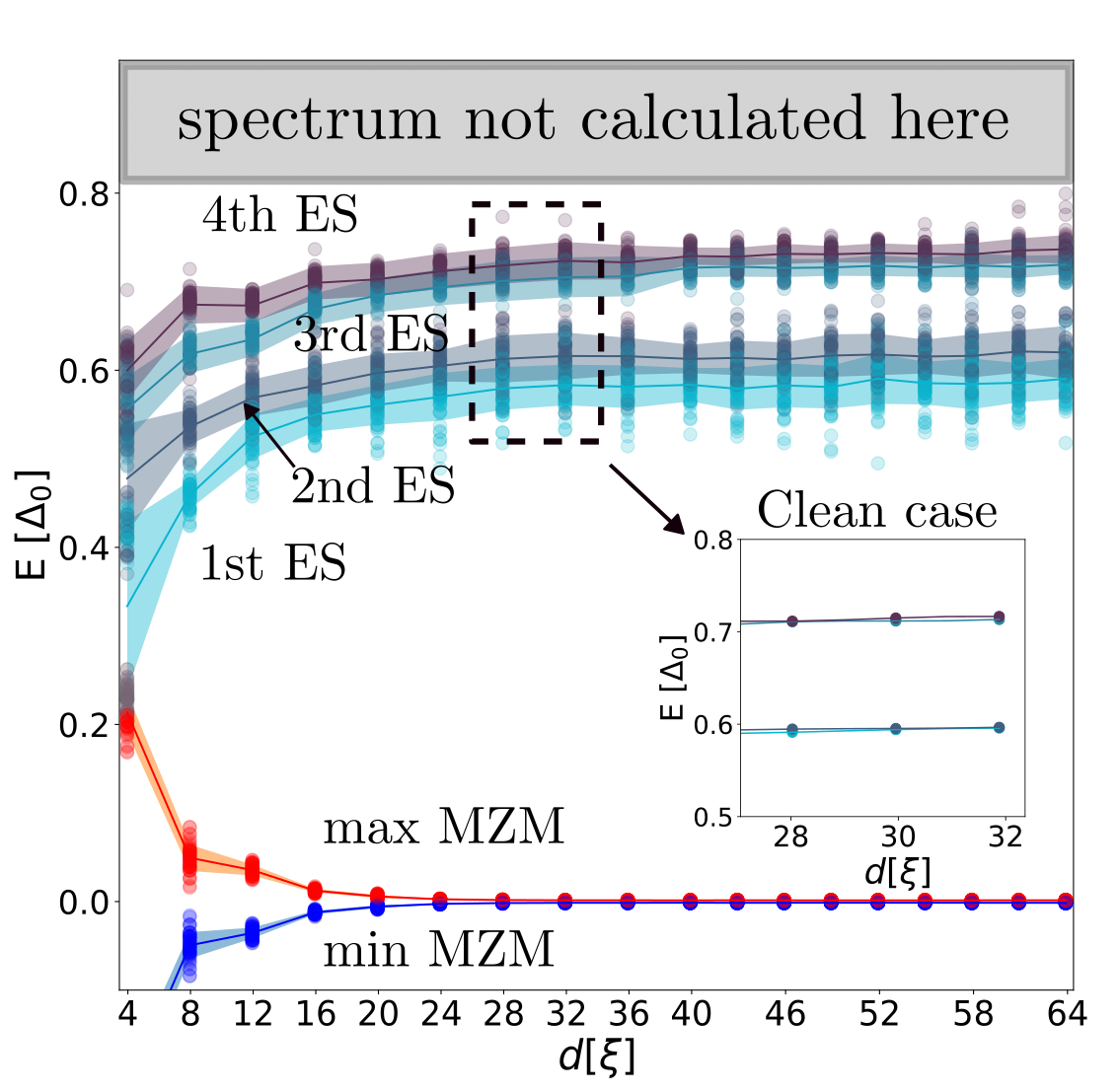}

\caption{
Eigenvalues vs distances with 50 realizations in the disordered system. 
The mean values and standard deviations are shown for each energy. 
The inset shows the spectrum in a clean system, illustrating the degeneracy of excited state pairs at large distances. This degeneracy arises from the  symmetry $\mathcal{R}_{\text{Z}_2}$  discussed below Eq.~(\ref{eq: two vortex}).
} \label{fig:Evsd}
\end{figure}

The two-vortex pairing term and the BdG Hamiltonian exhibit a $\text{Z}_2$ symmetry represented by $\mathcal{R}_{\text{Z}_2}=R(z,\pi)\tau_z \eta_z $. This operator is characterized by a $\pi$ rotation around the z-axis, with respect to the midpoint of the two vortices, taken here as the origin. Its action on a function $f(x,y,z)$ is given by $R(z,\pi)f(x,y,z)=f(-x,-y,z)$. 
The symmetry operator has eigenvalues $\pm1$. The manifestation of this symmetry results in the observation of double degeneracy in the system's spectrum, as evident in the inset of Fig.~\ref{fig:Evsd}. The operator  
$\mathcal{R}_{\text{Z}_2}$ commutes with the Hamiltonian~(\ref{eq:EM coupling}), establishing a selection rule that governs the allowed MW transitions within the system. According to the selection rule, transitions within the system can only occur between states that share the same eigenvalues of $\mathcal{R}_{\text{Z}_2}$. 
Since the PHS operator $\mathcal{P}=\tau_x K$ changes the eigenvalue of $\mathcal{R}_{\text{Z}_2}$, at least one of the transition matrix elements $q_{n,+\text{M}},q_{n,-\text{M}}$ vanishes. 

\begin{figure}
\centering
\includegraphics[width=1.0\columnwidth]{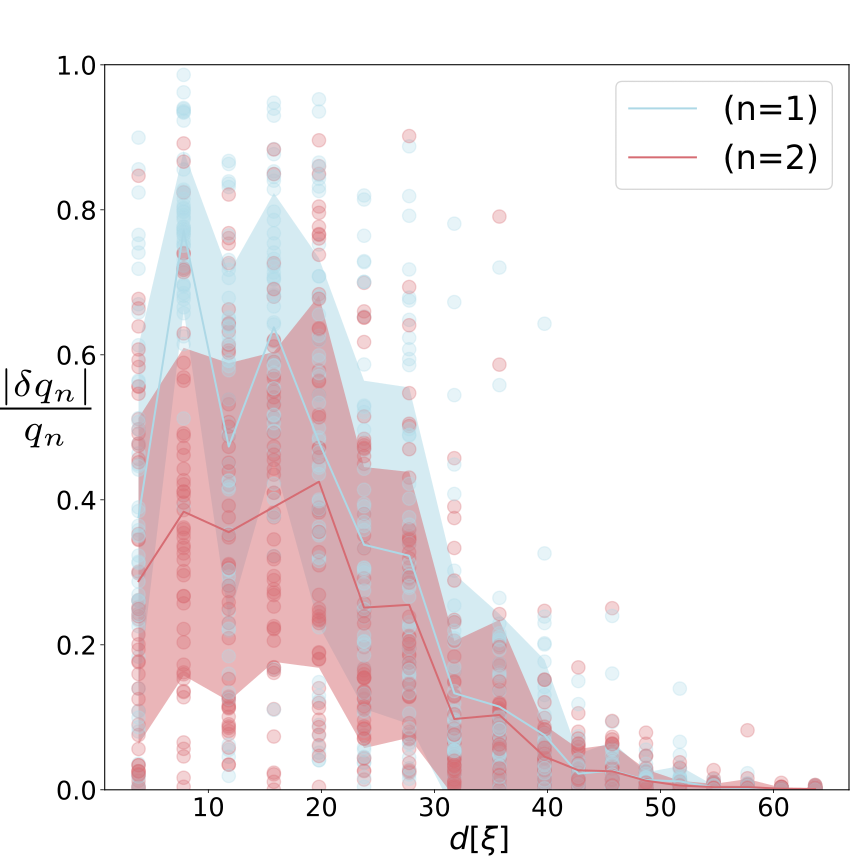}
\caption{
The ratio of parity-dependent charge difference $|\delta q_n|=|q_{n,+M}|-|q_{n,-M}|$ to the total  charge $q_n =|q_{n,+M}|+|q_{n,-M}|$ vs distance for  50 disorder realizations. 
The charges between MZMs and the 2 lowest excited states are shown with their mean values and standard deviations. The ratio in the clean system is always 1 due to the selection rule discussed below Eq.~(\ref{eq: two vortex}).
} \label{fig:ratio_charge}
\end{figure}

However, the presence of random disorder in realistic conditions disrupts the symmetry, resulting in the elimination of the double degeneracy in the spectrum (Fig.~\ref{fig:Evsd}). Consequently, this compromises the strict adherence to the selection rule. None of the transition matrix elements $q_{n,+\text{M}},q_{n,-\text{M}}$ are generally zero (Fig.~\ref{fig:ratio_charge}). Thus, in realistic experimental settings, the selection rule is not rigorously maintained.

\textbf{\textit{Numerical studies of two-vortex systems.~} } 
We employ a numerical approach to investigate a two-vortex system. To perform the numerical analysis, we discretize the Hamiltonian given by Eq.~(\ref{eq:3D SC}) and utilize the Kwant package~\cite{groth2014kwant} in Python to implement and solve the corresponding tight-binding model. The system under consideration is a cuboid with dimensions 500 nm $\cross$ 250 nm $\cross$ 25 nm (refer to Fig.~\ref{fig:system}a for an illustration), 
discretized with a lattice constant $a =5$ nm.

We utilize the results obtained in Ref.~\cite{SuppMat} to calculate the screened electric potential of two vortices~\cite{PhysRevLett.77.566}, which is then included in the real space version of the Hamiltonian in Eq.~(\ref{eq:3D SC}), similar to the way the chemical potential $\mu$ is incorporated. We take the screening length $\lambda_{\text{TF}}$ as one lattice constant. Our investigation encompasses both clean and disordered systems. To model the disorder, we introduce a position-dependent random potential into the Hamiltonian. The disorder potential follows a normal distribution, with the standard deviation of this distribution matching the gap $\Delta_0$. The spectrums and charge matrix elements acquired through numerical computations are depicted in Fig.~\ref{fig:Evsd} and Fig.~\ref{fig:ratio_charge}. 

\textbf{\textit{Discussion.~}}
We showed that a microwave coupling enables the parity readout of a non-local Majorana zero mode hosted in a vortex pair. 
We quantified the sensitivity of the readout by defining a critical cavity Q-factor $Q_c$, Eqs.~(\ref{eq:Qc1})-(\ref{eq:Qc2b}),  required of the resonant cavity coupled to the vortices. 
To estimate a typical value of $Q_c$, let us consider a resonant frequency 5 GHz (much below a typical superconducting gap $\Delta_0$) and effective capacitance   $1\times 10^{-12}$ F of a typical coplanar waveguide resonator~\cite{10.1063/1.3010859}. In our simulation, we find that the MZM energy $E_{\text{M}}$ for a system with a large vortex separation  $d=36 \xi$  can be neglected while the first excited state is approximately at $E_1 \approx 0.58\Delta_0 \gg \omega_c$ (see Fig.~\ref{fig:Evsd}), implying the system is in the detuning regime.
The relevant charge matrix elements are numerically estimated to be $q_1 \approx 0.009 e$ and $\delta q_1 \approx 0.002 e$, the ratio of which is shown in Fig.~\ref{fig:ratio_charge}. In this case, 
Eq.~(\ref{eq:Qc2a}) gives the  required critical cavity Q-factor $Q_c \sim 10^8$, which is close to state-of-the-art experimental conditions~\cite{8665969}. Below distance $d \approx 20\xi$, the system is still in the detuning regime of Eq.~(\ref{eq:Qc2a}). There, $Q_c \sim 10^6$, well within reach of the experiments. 

Our method offers a compelling approach to measuring the non-Abelian nature of Majorana zero modes. By employing two resonators to measure quantities $s_z=i\gamma_1 \gamma_2$ and $s_x=i\gamma_2 \gamma_3$, we can effectively measure two non-commuting parities of MZMs. Through monitoring these observables~\cite{PhysRevLett.113.247001,PhysRevLett.121.047001}, we can estimate quasi-particle poisoning time and MZM hybridization $E_{\text{M}}$.  Additionally, incorporating a third resonator to measure $s_y=i\gamma_1 \gamma_3$ and an ancillary pair of MZMs, would enable measurement-based braiding~\cite{PhysRevLett.101.010501,PhysRevApplied.12.054035,2017PhRvB..95w5305K} within timescales shorter than the quasi-particle poisoning time, when the total parity is conserved. Alternatively, braiding can be achieved through time-dependent control of MZM hybridization in a non-topologically protected manner~\cite{PhysRevLett.122.236803}. 
Thus, the resonator-based approach not only allows one to measure the essential  quasi-particle poisoning time but also enables one to demonstrate the non-Abelian characteristics of vortex-based MZMs,  thus holding  significant promise for advancing topological quantum computing and related technologies. 

\begin{acknowledgments} 
\textbf{\emph{Acknowledgments.~}} 
We thank Yong Chen, Valla Fatemi, Leonid Glazman, Mingi Kim, and Lingyuan Kong 
for valuable discussions. This work was initiated at Aspen Center for Physics, which is supported by National Science Foundation grant PHY-1607611. 
This material is based upon work supported by the Office of the Under Secretary of Defense for Research and Engineering under award number FA9550-22-1-0354. 
\end{acknowledgments}

\bibliography{refs}

\begin{thebibliography}{67}%
\makeatletter
\providecommand \@ifxundefined [1]{%
 \@ifx{#1\undefined}
}%
\providecommand \@ifnum [1]{%
 \ifnum #1\expandafter \@firstoftwo
 \else \expandafter \@secondoftwo
 \fi
}%
\providecommand \@ifx [1]{%
 \ifx #1\expandafter \@firstoftwo
 \else \expandafter \@secondoftwo
 \fi
}%
\providecommand \natexlab [1]{#1}%
\providecommand \enquote  [1]{``#1''}%
\providecommand \bibnamefont  [1]{#1}%
\providecommand \bibfnamefont [1]{#1}%
\providecommand \citenamefont [1]{#1}%
\providecommand \href@noop [0]{\@secondoftwo}%
\providecommand \href [0]{\begingroup \@sanitize@url \@href}%
\providecommand \@href[1]{\@@startlink{#1}\@@href}%
\providecommand \@@href[1]{\endgroup#1\@@endlink}%
\providecommand \@sanitize@url [0]{\catcode `\\12\catcode `\$12\catcode
  `\&12\catcode `\#12\catcode `\^12\catcode `\_12\catcode `\%12\relax}%
\providecommand \@@startlink[1]{}%
\providecommand \@@endlink[0]{}%
\providecommand \url  [0]{\begingroup\@sanitize@url \@url }%
\providecommand \@url [1]{\endgroup\@href {#1}{\urlprefix }}%
\providecommand \urlprefix  [0]{URL }%
\providecommand \Eprint [0]{\href }%
\providecommand \doibase [0]{http://dx.doi.org/}%
\providecommand \selectlanguage [0]{\@gobble}%
\providecommand \bibinfo  [0]{\@secondoftwo}%
\providecommand \bibfield  [0]{\@secondoftwo}%
\providecommand \translation [1]{[#1]}%
\providecommand \BibitemOpen [0]{}%
\providecommand \bibitemStop [0]{}%
\providecommand \bibitemNoStop [0]{.\EOS\space}%
\providecommand \EOS [0]{\spacefactor3000\relax}%
\providecommand \BibitemShut  [1]{\csname bibitem#1\endcsname}%
\let\auto@bib@innerbib\@empty
\bibitem [{\citenamefont {{Volovik}}(1999)}]{Volovik1999}%
  \BibitemOpen
  \bibfield  {author} {\bibinfo {author} {\bibfnamefont {G.~E.}\ \bibnamefont
  {{Volovik}}},\ }\bibfield  {title} {\enquote {\bibinfo {title} {{Fermion zero
  modes on vortices in chiral superconductors}},}\ }\href {\doibase
  10.1134/1.568223} {\bibfield  {journal} {\bibinfo  {journal} {Soviet Journal
  of Experimental and Theoretical Physics Letters}\ }\textbf {\bibinfo {volume}
  {70}},\ \bibinfo {pages} {609--614} (\bibinfo {year} {1999})}\BibitemShut
  {NoStop}%
\bibitem [{\citenamefont {{Read}}\ and\ \citenamefont
  {{Green}}(2000)}]{2000PhRvB..6110267R}%
  \BibitemOpen
  \bibfield  {author} {\bibinfo {author} {\bibfnamefont {N.}~\bibnamefont
  {{Read}}}\ and\ \bibinfo {author} {\bibfnamefont {Dmitry}\ \bibnamefont
  {{Green}}},\ }\bibfield  {title} {\enquote {\bibinfo {title} {{Paired states
  of fermions in two dimensions with breaking of parity and time-reversal
  symmetries and the fractional quantum Hall effect}},}\ }\href {\doibase
  10.1103/PhysRevB.61.10267} {\bibfield  {journal} {\bibinfo  {journal}
  {Physical Review B}\ }\textbf {\bibinfo {volume} {61}},\ \bibinfo {pages}
  {10267--10297} (\bibinfo {year} {2000})},\ \Eprint
  {http://arxiv.org/abs/cond-mat/9906453} {arXiv:cond-mat/9906453
  [cond-mat.mes-hall]} \BibitemShut {NoStop}%
\bibitem [{\citenamefont {{Ivanov}}(2001)}]{2001PhRvL..86..268I}%
  \BibitemOpen
  \bibfield  {author} {\bibinfo {author} {\bibfnamefont {D.~A.}\ \bibnamefont
  {{Ivanov}}},\ }\bibfield  {title} {\enquote {\bibinfo {title} {{Non-Abelian
  Statistics of Half-Quantum Vortices in p-Wave Superconductors}},}\ }\href
  {\doibase 10.1103/PhysRevLett.86.268} {\bibfield  {journal} {\bibinfo
  {journal} {Phys. Rev. Lett.}\ }\textbf {\bibinfo {volume} {86}},\ \bibinfo
  {pages} {268--271} (\bibinfo {year} {2001})}\BibitemShut {NoStop}%
\bibitem [{\citenamefont {{Fu}}\ and\ \citenamefont
  {{Kane}}(2008)}]{2008PhRvL.100i6407F}%
  \BibitemOpen
  \bibfield  {author} {\bibinfo {author} {\bibfnamefont {L.}~\bibnamefont
  {{Fu}}}\ and\ \bibinfo {author} {\bibfnamefont {C.~L.}\ \bibnamefont
  {{Kane}}},\ }\bibfield  {title} {\enquote {\bibinfo {title} {{Superconducting
  Proximity Effect and Majorana Fermions at the Surface of a Topological
  Insulator}},}\ }\href {\doibase 10.1103/PhysRevLett.100.096407} {\bibfield
  {journal} {\bibinfo  {journal} {Phys. Rev. Lett.}\ }\textbf {\bibinfo
  {volume} {100}},\ \bibinfo {eid} {096407} (\bibinfo {year}
  {2008})}\BibitemShut {NoStop}%
\bibitem [{\citenamefont {Sau}\ \emph {et~al.}(2010)\citenamefont {Sau},
  \citenamefont {Lutchyn}, \citenamefont {Tewari},\ and\ \citenamefont
  {Das~Sarma}}]{2010PhRvL.104d0502S}%
  \BibitemOpen
  \bibfield  {author} {\bibinfo {author} {\bibfnamefont {Jay~D.}\ \bibnamefont
  {Sau}}, \bibinfo {author} {\bibfnamefont {Roman~M.}\ \bibnamefont {Lutchyn}},
  \bibinfo {author} {\bibfnamefont {Sumanta}\ \bibnamefont {Tewari}}, \ and\
  \bibinfo {author} {\bibfnamefont {S.}~\bibnamefont {Das~Sarma}},\ }\bibfield
  {title} {\enquote {\bibinfo {title} {{Generic New Platform for Topological
  Quantum Computation Using Semiconductor Heterostructures}},}\ }\href
  {\doibase 10.1103/PhysRevLett.104.040502} {\bibfield  {journal} {\bibinfo
  {journal} {Phys. Rev. Lett.}\ }\textbf {\bibinfo {volume} {104}},\ \bibinfo
  {pages} {040502} (\bibinfo {year} {2010})}\BibitemShut {NoStop}%
\bibitem [{\citenamefont {Chen}\ \emph {et~al.}(2018)\citenamefont {Chen},
  \citenamefont {Chen}, \citenamefont {Yang}, \citenamefont {Du}, \citenamefont
  {Zhu}, \citenamefont {Wang},\ and\ \citenamefont {Wen}}]{chen2018discrete}%
  \BibitemOpen
  \bibfield  {author} {\bibinfo {author} {\bibfnamefont {Mingyang}\
  \bibnamefont {Chen}}, \bibinfo {author} {\bibfnamefont {Xiaoyu}\ \bibnamefont
  {Chen}}, \bibinfo {author} {\bibfnamefont {Huan}\ \bibnamefont {Yang}},
  \bibinfo {author} {\bibfnamefont {Zengyi}\ \bibnamefont {Du}}, \bibinfo
  {author} {\bibfnamefont {Xiyu}\ \bibnamefont {Zhu}}, \bibinfo {author}
  {\bibfnamefont {Enyu}\ \bibnamefont {Wang}}, \ and\ \bibinfo {author}
  {\bibfnamefont {Hai-Hu}\ \bibnamefont {Wen}},\ }\bibfield  {title} {\enquote
  {\bibinfo {title} {{Discrete energy levels of Caroli-de Gennes-Matricon
  states in quantum limit in FeTe$_{0.55}$Se$_{0.45}$}},}\ }\href {\doibase
  10.1038/s41467-018-03404-8} {\bibfield  {journal} {\bibinfo  {journal}
  {Nature communications}\ }\textbf {\bibinfo {volume} {9}},\ \bibinfo {pages}
  {1--7} (\bibinfo {year} {2018})}\BibitemShut {NoStop}%
\bibitem [{\citenamefont {Wang}\ \emph {et~al.}(2018)\citenamefont {Wang},
  \citenamefont {Kong}, \citenamefont {Fan}, \citenamefont {Chen},
  \citenamefont {Zhu}, \citenamefont {Liu}, \citenamefont {Cao}, \citenamefont
  {Sun}, \citenamefont {Du}, \citenamefont {Schneeloch}, \citenamefont {Zhong},
  \citenamefont {Gu}, \citenamefont {Fu}, \citenamefont {Ding},\ and\
  \citenamefont {Gao}}]{doi:10.1126/science.aao1797}%
  \BibitemOpen
  \bibfield  {author} {\bibinfo {author} {\bibfnamefont {Dongfei}\ \bibnamefont
  {Wang}}, \bibinfo {author} {\bibfnamefont {Lingyuan}\ \bibnamefont {Kong}},
  \bibinfo {author} {\bibfnamefont {Peng}\ \bibnamefont {Fan}}, \bibinfo
  {author} {\bibfnamefont {Hui}\ \bibnamefont {Chen}}, \bibinfo {author}
  {\bibfnamefont {Shiyu}\ \bibnamefont {Zhu}}, \bibinfo {author} {\bibfnamefont
  {Wenyao}\ \bibnamefont {Liu}}, \bibinfo {author} {\bibfnamefont
  {Lu}~\bibnamefont {Cao}}, \bibinfo {author} {\bibfnamefont {Yujie}\
  \bibnamefont {Sun}}, \bibinfo {author} {\bibfnamefont {Shixuan}\ \bibnamefont
  {Du}}, \bibinfo {author} {\bibfnamefont {John}\ \bibnamefont {Schneeloch}},
  \bibinfo {author} {\bibfnamefont {Ruidan}\ \bibnamefont {Zhong}}, \bibinfo
  {author} {\bibfnamefont {Genda}\ \bibnamefont {Gu}}, \bibinfo {author}
  {\bibfnamefont {Liang}\ \bibnamefont {Fu}}, \bibinfo {author} {\bibfnamefont
  {Hong}\ \bibnamefont {Ding}}, \ and\ \bibinfo {author} {\bibfnamefont
  {Hong-Jun}\ \bibnamefont {Gao}},\ }\bibfield  {title} {\enquote {\bibinfo
  {title} {{Evidence for Majorana bound states in an iron-based
  superconductor}},}\ }\href {\doibase 10.1126/science.aao1797} {\bibfield
  {journal} {\bibinfo  {journal} {Science}\ }\textbf {\bibinfo {volume}
  {362}},\ \bibinfo {pages} {333--335} (\bibinfo {year} {2018})}\BibitemShut
  {NoStop}%
\bibitem [{\citenamefont {Liu}\ \emph {et~al.}(2018)\citenamefont {Liu},
  \citenamefont {Chen}, \citenamefont {Zhang}, \citenamefont {Peng},
  \citenamefont {Yan}, \citenamefont {Wen}, \citenamefont {Lou}, \citenamefont
  {Huang}, \citenamefont {Tian}, \citenamefont {Dong}, \citenamefont {Wang},
  \citenamefont {Bao}, \citenamefont {Wang}, \citenamefont {Yin}, \citenamefont
  {Zhao},\ and\ \citenamefont {Feng}}]{PhysRevX.8.041056}%
  \BibitemOpen
  \bibfield  {author} {\bibinfo {author} {\bibfnamefont {Qin}\ \bibnamefont
  {Liu}}, \bibinfo {author} {\bibfnamefont {Chen}\ \bibnamefont {Chen}},
  \bibinfo {author} {\bibfnamefont {Tong}\ \bibnamefont {Zhang}}, \bibinfo
  {author} {\bibfnamefont {Rui}\ \bibnamefont {Peng}}, \bibinfo {author}
  {\bibfnamefont {Ya-Jun}\ \bibnamefont {Yan}}, \bibinfo {author}
  {\bibfnamefont {Chen-Hao-Ping}\ \bibnamefont {Wen}}, \bibinfo {author}
  {\bibfnamefont {Xia}\ \bibnamefont {Lou}}, \bibinfo {author} {\bibfnamefont
  {Yu-Long}\ \bibnamefont {Huang}}, \bibinfo {author} {\bibfnamefont
  {Jin-Peng}\ \bibnamefont {Tian}}, \bibinfo {author} {\bibfnamefont {Xiao-Li}\
  \bibnamefont {Dong}}, \bibinfo {author} {\bibfnamefont {Guang-Wei}\
  \bibnamefont {Wang}}, \bibinfo {author} {\bibfnamefont {Wei-Cheng}\
  \bibnamefont {Bao}}, \bibinfo {author} {\bibfnamefont {Qiang-Hua}\
  \bibnamefont {Wang}}, \bibinfo {author} {\bibfnamefont {Zhi-Ping}\
  \bibnamefont {Yin}}, \bibinfo {author} {\bibfnamefont {Zhong-Xian}\
  \bibnamefont {Zhao}}, \ and\ \bibinfo {author} {\bibfnamefont {Dong-Lai}\
  \bibnamefont {Feng}},\ }\bibfield  {title} {\enquote {\bibinfo {title}
  {{Robust and Clean Majorana Zero Mode in the Vortex Core of High-Temperature
  Superconductor
  $\mathbf{(}{\mathrm{Li}}_{0.84}{\mathrm{Fe}}_{0.16}\mathbf{)}\mathrm{OHFeSe}$}},}\
  }\href {\doibase 10.1103/PhysRevX.8.041056} {\bibfield  {journal} {\bibinfo
  {journal} {Phys. Rev. X}\ }\textbf {\bibinfo {volume} {8}},\ \bibinfo {pages}
  {041056} (\bibinfo {year} {2018})}\BibitemShut {NoStop}%
\bibitem [{\citenamefont {Chen}\ \emph {et~al.}(2019)\citenamefont {Chen},
  \citenamefont {Liu}, \citenamefont {Zhang}, \citenamefont {Li}, \citenamefont
  {Shen}, \citenamefont {Dong}, \citenamefont {Zhao}, \citenamefont {Zhang},\
  and\ \citenamefont {Feng}}]{Chen_2019}%
  \BibitemOpen
  \bibfield  {author} {\bibinfo {author} {\bibfnamefont {C.}~\bibnamefont
  {Chen}}, \bibinfo {author} {\bibfnamefont {Q.}~\bibnamefont {Liu}}, \bibinfo
  {author} {\bibfnamefont {T.~Z.}\ \bibnamefont {Zhang}}, \bibinfo {author}
  {\bibfnamefont {D.}~\bibnamefont {Li}}, \bibinfo {author} {\bibfnamefont
  {P.~P.}\ \bibnamefont {Shen}}, \bibinfo {author} {\bibfnamefont {X.~L.}\
  \bibnamefont {Dong}}, \bibinfo {author} {\bibfnamefont {Z.-X.}\ \bibnamefont
  {Zhao}}, \bibinfo {author} {\bibfnamefont {T.}~\bibnamefont {Zhang}}, \ and\
  \bibinfo {author} {\bibfnamefont {D.~L.}\ \bibnamefont {Feng}},\ }\bibfield
  {title} {\enquote {\bibinfo {title} {{Quantized Conductance of Majorana Zero
  Mode in the Vortex of the Topological Superconductor
  $\mathbf{(}{\mathrm{Li}}_{0.84}{\mathrm{Fe}}_{0.16}\mathbf{)}\mathrm{OHFeSe}$}},}\
  }\href {\doibase 10.1088/0256-307X/36/5/057403} {\bibfield  {journal}
  {\bibinfo  {journal} {Chinese Physics Letters}\ }\textbf {\bibinfo {volume}
  {36}},\ \bibinfo {pages} {057403} (\bibinfo {year} {2019})}\BibitemShut
  {NoStop}%
\bibitem [{\citenamefont {Machida}\ \emph {et~al.}(2019)\citenamefont
  {Machida}, \citenamefont {Sun}, \citenamefont {Pyon}, \citenamefont {Takeda},
  \citenamefont {Kohsaka}, \citenamefont {Hanaguri}, \citenamefont {Sasagawa},\
  and\ \citenamefont {Tamegai}}]{Machida2019}%
  \BibitemOpen
  \bibfield  {author} {\bibinfo {author} {\bibfnamefont {T.}~\bibnamefont
  {Machida}}, \bibinfo {author} {\bibfnamefont {Y.}~\bibnamefont {Sun}},
  \bibinfo {author} {\bibfnamefont {S.}~\bibnamefont {Pyon}}, \bibinfo {author}
  {\bibfnamefont {S.}~\bibnamefont {Takeda}}, \bibinfo {author} {\bibfnamefont
  {Y.}~\bibnamefont {Kohsaka}}, \bibinfo {author} {\bibfnamefont
  {T.}~\bibnamefont {Hanaguri}}, \bibinfo {author} {\bibfnamefont
  {T.}~\bibnamefont {Sasagawa}}, \ and\ \bibinfo {author} {\bibfnamefont
  {T.}~\bibnamefont {Tamegai}},\ }\bibfield  {title} {\enquote {\bibinfo
  {title} {{Zero-energy vortex bound state in the superconducting topological
  surface state of Fe(Se,Te)}},}\ }\href {\doibase 10.1038/s41563-019-0397-1}
  {\bibfield  {journal} {\bibinfo  {journal} {Nature Materials}\ }\textbf
  {\bibinfo {volume} {18}},\ \bibinfo {pages} {811--815} (\bibinfo {year}
  {2019})}\BibitemShut {NoStop}%
\bibitem [{\citenamefont {{Kong}}\ \emph {et~al.}(2019)\citenamefont {{Kong}},
  \citenamefont {{Zhu}}, \citenamefont {{Papaj}}, \citenamefont {{Chen}},
  \citenamefont {{Cao}}, \citenamefont {{Isobe}}, \citenamefont {{Xing}},
  \citenamefont {{Liu}}, \citenamefont {{Wang}}, \citenamefont {{Fan}},
  \citenamefont {{Sun}}, \citenamefont {{Du}}, \citenamefont {{Schneeloch}},
  \citenamefont {{Zhong}}, \citenamefont {{Gu}}, \citenamefont {{Fu}},
  \citenamefont {{Gao}},\ and\ \citenamefont {{Ding}}}]{2019NatPh..15.1181K}%
  \BibitemOpen
  \bibfield  {author} {\bibinfo {author} {\bibfnamefont {Lingyuan}\
  \bibnamefont {{Kong}}}, \bibinfo {author} {\bibfnamefont {Shiyu}\
  \bibnamefont {{Zhu}}}, \bibinfo {author} {\bibfnamefont {Micha{\l}}\
  \bibnamefont {{Papaj}}}, \bibinfo {author} {\bibfnamefont {Hui}\ \bibnamefont
  {{Chen}}}, \bibinfo {author} {\bibfnamefont {Lu}~\bibnamefont {{Cao}}},
  \bibinfo {author} {\bibfnamefont {Hiroki}\ \bibnamefont {{Isobe}}}, \bibinfo
  {author} {\bibfnamefont {Yuqing}\ \bibnamefont {{Xing}}}, \bibinfo {author}
  {\bibfnamefont {Wenyao}\ \bibnamefont {{Liu}}}, \bibinfo {author}
  {\bibfnamefont {Dongfei}\ \bibnamefont {{Wang}}}, \bibinfo {author}
  {\bibfnamefont {Peng}\ \bibnamefont {{Fan}}}, \bibinfo {author}
  {\bibfnamefont {Yujie}\ \bibnamefont {{Sun}}}, \bibinfo {author}
  {\bibfnamefont {Shixuan}\ \bibnamefont {{Du}}}, \bibinfo {author}
  {\bibfnamefont {John}\ \bibnamefont {{Schneeloch}}}, \bibinfo {author}
  {\bibfnamefont {Ruidan}\ \bibnamefont {{Zhong}}}, \bibinfo {author}
  {\bibfnamefont {Genda}\ \bibnamefont {{Gu}}}, \bibinfo {author}
  {\bibfnamefont {Liang}\ \bibnamefont {{Fu}}}, \bibinfo {author}
  {\bibfnamefont {Hong-Jun}\ \bibnamefont {{Gao}}}, \ and\ \bibinfo {author}
  {\bibfnamefont {Hong}\ \bibnamefont {{Ding}}},\ }\bibfield  {title} {\enquote
  {\bibinfo {title} {{Half-integer level shift of vortex bound states in an
  iron-based superconductor}},}\ }\href {\doibase 10.1038/s41567-019-0630-5}
  {\bibfield  {journal} {\bibinfo  {journal} {Nature Physics}\ }\textbf
  {\bibinfo {volume} {15}},\ \bibinfo {pages} {1181--1187} (\bibinfo {year}
  {2019})}\BibitemShut {NoStop}%
\bibitem [{\citenamefont {Zhu}\ \emph {et~al.}(2020)\citenamefont {Zhu},
  \citenamefont {Kong}, \citenamefont {Cao}, \citenamefont {Chen},
  \citenamefont {Papaj}, \citenamefont {Du}, \citenamefont {Xing},
  \citenamefont {Liu}, \citenamefont {Wang}, \citenamefont {Shen},
  \citenamefont {Yang}, \citenamefont {Schneeloch}, \citenamefont {Zhong},
  \citenamefont {Gu}, \citenamefont {Fu}, \citenamefont {Zhang}, \citenamefont
  {Ding},\ and\ \citenamefont {Gao}}]{doi:10.1126/science.aax0274}%
  \BibitemOpen
  \bibfield  {author} {\bibinfo {author} {\bibfnamefont {Shiyu}\ \bibnamefont
  {Zhu}}, \bibinfo {author} {\bibfnamefont {Lingyuan}\ \bibnamefont {Kong}},
  \bibinfo {author} {\bibfnamefont {Lu}~\bibnamefont {Cao}}, \bibinfo {author}
  {\bibfnamefont {Hui}\ \bibnamefont {Chen}}, \bibinfo {author} {\bibfnamefont
  {Michał}\ \bibnamefont {Papaj}}, \bibinfo {author} {\bibfnamefont {Shixuan}\
  \bibnamefont {Du}}, \bibinfo {author} {\bibfnamefont {Yuqing}\ \bibnamefont
  {Xing}}, \bibinfo {author} {\bibfnamefont {Wenyao}\ \bibnamefont {Liu}},
  \bibinfo {author} {\bibfnamefont {Dongfei}\ \bibnamefont {Wang}}, \bibinfo
  {author} {\bibfnamefont {Chengmin}\ \bibnamefont {Shen}}, \bibinfo {author}
  {\bibfnamefont {Fazhi}\ \bibnamefont {Yang}}, \bibinfo {author}
  {\bibfnamefont {John}\ \bibnamefont {Schneeloch}}, \bibinfo {author}
  {\bibfnamefont {Ruidan}\ \bibnamefont {Zhong}}, \bibinfo {author}
  {\bibfnamefont {Genda}\ \bibnamefont {Gu}}, \bibinfo {author} {\bibfnamefont
  {Liang}\ \bibnamefont {Fu}}, \bibinfo {author} {\bibfnamefont {Yu-Yang}\
  \bibnamefont {Zhang}}, \bibinfo {author} {\bibfnamefont {Hong}\ \bibnamefont
  {Ding}}, \ and\ \bibinfo {author} {\bibfnamefont {Hong-Jun}\ \bibnamefont
  {Gao}},\ }\bibfield  {title} {\enquote {\bibinfo {title} {Nearly quantized
  conductance plateau of vortex zero mode in an iron-based superconductor},}\
  }\href {\doibase 10.1126/science.aax0274} {\bibfield  {journal} {\bibinfo
  {journal} {Science}\ }\textbf {\bibinfo {volume} {367}},\ \bibinfo {pages}
  {189--192} (\bibinfo {year} {2020})}\BibitemShut {NoStop}%
\bibitem [{\citenamefont {Liu}\ \emph {et~al.}(2020)\citenamefont {Liu},
  \citenamefont {Cao}, \citenamefont {Zhu}, \citenamefont {Kong}, \citenamefont
  {Wang}, \citenamefont {Papaj}, \citenamefont {Zhang}, \citenamefont {Liu},
  \citenamefont {Chen}, \citenamefont {Li}, \citenamefont {Yang}, \citenamefont
  {Kondo}, \citenamefont {Du}, \citenamefont {Cao}, \citenamefont {Shin},
  \citenamefont {Fu}, \citenamefont {Yin}, \citenamefont {Gao},\ and\
  \citenamefont {Ding}}]{Liu2020}%
  \BibitemOpen
  \bibfield  {author} {\bibinfo {author} {\bibfnamefont {Wenyao}\ \bibnamefont
  {Liu}}, \bibinfo {author} {\bibfnamefont {Lu}~\bibnamefont {Cao}}, \bibinfo
  {author} {\bibfnamefont {Shiyu}\ \bibnamefont {Zhu}}, \bibinfo {author}
  {\bibfnamefont {Lingyuan}\ \bibnamefont {Kong}}, \bibinfo {author}
  {\bibfnamefont {Guangwei}\ \bibnamefont {Wang}}, \bibinfo {author}
  {\bibfnamefont {Micha{\l}}\ \bibnamefont {Papaj}}, \bibinfo {author}
  {\bibfnamefont {Peng}\ \bibnamefont {Zhang}}, \bibinfo {author}
  {\bibfnamefont {Ya-Bin}\ \bibnamefont {Liu}}, \bibinfo {author}
  {\bibfnamefont {Hui}\ \bibnamefont {Chen}}, \bibinfo {author} {\bibfnamefont
  {Geng}\ \bibnamefont {Li}}, \bibinfo {author} {\bibfnamefont {Fazhi}\
  \bibnamefont {Yang}}, \bibinfo {author} {\bibfnamefont {Takeshi}\
  \bibnamefont {Kondo}}, \bibinfo {author} {\bibfnamefont {Shixuan}\
  \bibnamefont {Du}}, \bibinfo {author} {\bibfnamefont {Guang-Han}\
  \bibnamefont {Cao}}, \bibinfo {author} {\bibfnamefont {Shik}\ \bibnamefont
  {Shin}}, \bibinfo {author} {\bibfnamefont {Liang}\ \bibnamefont {Fu}},
  \bibinfo {author} {\bibfnamefont {Zhiping}\ \bibnamefont {Yin}}, \bibinfo
  {author} {\bibfnamefont {Hong-Jun}\ \bibnamefont {Gao}}, \ and\ \bibinfo
  {author} {\bibfnamefont {Hong}\ \bibnamefont {Ding}},\ }\bibfield  {title}
  {\enquote {\bibinfo {title} {{A new Majorana platform in an Fe-As bilayer
  superconductor}},}\ }\href {\doibase 10.1038/s41467-020-19487-1} {\bibfield
  {journal} {\bibinfo  {journal} {Nature Communications}\ }\textbf {\bibinfo
  {volume} {11}},\ \bibinfo {pages} {5688} (\bibinfo {year}
  {2020})}\BibitemShut {NoStop}%
\bibitem [{\citenamefont {Kong}\ \emph {et~al.}(2021)\citenamefont {Kong},
  \citenamefont {Cao}, \citenamefont {Zhu}, \citenamefont {Papaj},
  \citenamefont {Dai}, \citenamefont {Li}, \citenamefont {Fan}, \citenamefont
  {Liu}, \citenamefont {Yang}, \citenamefont {Wang}, \citenamefont {Du},
  \citenamefont {Jin}, \citenamefont {Fu}, \citenamefont {Gao},\ and\
  \citenamefont {Ding}}]{Kong2021}%
  \BibitemOpen
  \bibfield  {author} {\bibinfo {author} {\bibfnamefont {Lingyuan}\
  \bibnamefont {Kong}}, \bibinfo {author} {\bibfnamefont {Lu}~\bibnamefont
  {Cao}}, \bibinfo {author} {\bibfnamefont {Shiyu}\ \bibnamefont {Zhu}},
  \bibinfo {author} {\bibfnamefont {Micha{\l}}\ \bibnamefont {Papaj}}, \bibinfo
  {author} {\bibfnamefont {Guangyang}\ \bibnamefont {Dai}}, \bibinfo {author}
  {\bibfnamefont {Geng}\ \bibnamefont {Li}}, \bibinfo {author} {\bibfnamefont
  {Peng}\ \bibnamefont {Fan}}, \bibinfo {author} {\bibfnamefont {Wenyao}\
  \bibnamefont {Liu}}, \bibinfo {author} {\bibfnamefont {Fazhi}\ \bibnamefont
  {Yang}}, \bibinfo {author} {\bibfnamefont {Xiancheng}\ \bibnamefont {Wang}},
  \bibinfo {author} {\bibfnamefont {Shixuan}\ \bibnamefont {Du}}, \bibinfo
  {author} {\bibfnamefont {Changqing}\ \bibnamefont {Jin}}, \bibinfo {author}
  {\bibfnamefont {Liang}\ \bibnamefont {Fu}}, \bibinfo {author} {\bibfnamefont
  {Hong-Jun}\ \bibnamefont {Gao}}, \ and\ \bibinfo {author} {\bibfnamefont
  {Hong}\ \bibnamefont {Ding}},\ }\bibfield  {title} {\enquote {\bibinfo
  {title} {{Majorana zero modes in impurity-assisted vortex of LiFeAs
  superconductor}},}\ }\href {\doibase 10.1038/s41467-021-24372-6} {\bibfield
  {journal} {\bibinfo  {journal} {Nature Communications}\ }\textbf {\bibinfo
  {volume} {12}},\ \bibinfo {pages} {4146} (\bibinfo {year}
  {2021})}\BibitemShut {NoStop}%
\bibitem [{\citenamefont {Shi}\ \emph {et~al.}(2017)\citenamefont {Shi},
  \citenamefont {Han}, \citenamefont {Richard}, \citenamefont {Wu},
  \citenamefont {Peng}, \citenamefont {Qian}, \citenamefont {Wang},
  \citenamefont {Hu}, \citenamefont {Sun},\ and\ \citenamefont
  {Ding}}]{SHI2017503}%
  \BibitemOpen
  \bibfield  {author} {\bibinfo {author} {\bibfnamefont {Xun}\ \bibnamefont
  {Shi}}, \bibinfo {author} {\bibfnamefont {Zhi-Qing}\ \bibnamefont {Han}},
  \bibinfo {author} {\bibfnamefont {Pierre}\ \bibnamefont {Richard}}, \bibinfo
  {author} {\bibfnamefont {Xian-Xin}\ \bibnamefont {Wu}}, \bibinfo {author}
  {\bibfnamefont {Xi-Liang}\ \bibnamefont {Peng}}, \bibinfo {author}
  {\bibfnamefont {Tian}\ \bibnamefont {Qian}}, \bibinfo {author} {\bibfnamefont
  {Shan-Cai}\ \bibnamefont {Wang}}, \bibinfo {author} {\bibfnamefont
  {Jiang-Ping}\ \bibnamefont {Hu}}, \bibinfo {author} {\bibfnamefont {Yu-Jie}\
  \bibnamefont {Sun}}, \ and\ \bibinfo {author} {\bibfnamefont {Hong}\
  \bibnamefont {Ding}},\ }\bibfield  {title} {\enquote {\bibinfo {title}
  {{FeTe$_{1-x}$Se$_x$ monolayer films: towards the realization of
  high-temperature connate topological superconductivity}},}\ }\href {\doibase
  https://doi.org/10.1016/j.scib.2017.03.010} {\bibfield  {journal} {\bibinfo
  {journal} {Science Bulletin}\ }\textbf {\bibinfo {volume} {62}},\ \bibinfo
  {pages} {503--507} (\bibinfo {year} {2017})}\BibitemShut {NoStop}%
\bibitem [{\citenamefont {Zhang}\ \emph {et~al.}(2018)\citenamefont {Zhang},
  \citenamefont {Yaji}, \citenamefont {Hashimoto}, \citenamefont {Ota},
  \citenamefont {Kondo}, \citenamefont {Okazaki}, \citenamefont {Wang},
  \citenamefont {Wen}, \citenamefont {Gu}, \citenamefont {Ding},\ and\
  \citenamefont {Shin}}]{doi:10.1126/science.aan4596}%
  \BibitemOpen
  \bibfield  {author} {\bibinfo {author} {\bibfnamefont {Peng}\ \bibnamefont
  {Zhang}}, \bibinfo {author} {\bibfnamefont {Koichiro}\ \bibnamefont {Yaji}},
  \bibinfo {author} {\bibfnamefont {Takahiro}\ \bibnamefont {Hashimoto}},
  \bibinfo {author} {\bibfnamefont {Yuichi}\ \bibnamefont {Ota}}, \bibinfo
  {author} {\bibfnamefont {Takeshi}\ \bibnamefont {Kondo}}, \bibinfo {author}
  {\bibfnamefont {Kozo}\ \bibnamefont {Okazaki}}, \bibinfo {author}
  {\bibfnamefont {Zhijun}\ \bibnamefont {Wang}}, \bibinfo {author}
  {\bibfnamefont {Jinsheng}\ \bibnamefont {Wen}}, \bibinfo {author}
  {\bibfnamefont {G.~D.}\ \bibnamefont {Gu}}, \bibinfo {author} {\bibfnamefont
  {Hong}\ \bibnamefont {Ding}}, \ and\ \bibinfo {author} {\bibfnamefont {Shik}\
  \bibnamefont {Shin}},\ }\bibfield  {title} {\enquote {\bibinfo {title}
  {Observation of topological superconductivity on the surface of an iron-based
  superconductor},}\ }\href {\doibase 10.1126/science.aan4596} {\bibfield
  {journal} {\bibinfo  {journal} {Science}\ }\textbf {\bibinfo {volume}
  {360}},\ \bibinfo {pages} {182--186} (\bibinfo {year} {2018})}\BibitemShut
  {NoStop}%
\bibitem [{\citenamefont {Zhang}\ \emph
  {et~al.}(2019{\natexlab{a}})\citenamefont {Zhang} \emph
  {et~al.}}]{Zhang2019}%
  \BibitemOpen
  \bibfield  {author} {\bibinfo {author} {\bibfnamefont {Peng}\ \bibnamefont
  {Zhang}} \emph {et~al.},\ }\bibfield  {title} {\enquote {\bibinfo {title}
  {Multiple topological states in iron-based superconductors},}\ }\href
  {\doibase 10.1038/s41567-018-0280-z} {\bibfield  {journal} {\bibinfo
  {journal} {Nature Physics}\ }\textbf {\bibinfo {volume} {15}},\ \bibinfo
  {pages} {41--47} (\bibinfo {year} {2019}{\natexlab{a}})}\BibitemShut
  {NoStop}%
\bibitem [{\citenamefont {Peng}\ \emph {et~al.}(2019)\citenamefont {Peng},
  \citenamefont {Li}, \citenamefont {Wu}, \citenamefont {Deng}, \citenamefont
  {Shi}, \citenamefont {Fan}, \citenamefont {Li}, \citenamefont {Huang},
  \citenamefont {Qian}, \citenamefont {Richard}, \citenamefont {Hu},
  \citenamefont {Pan}, \citenamefont {Mao}, \citenamefont {Sun},\ and\
  \citenamefont {Ding}}]{PhysRevB.100.155134}%
  \BibitemOpen
  \bibfield  {author} {\bibinfo {author} {\bibfnamefont {X.-L.}\ \bibnamefont
  {Peng}}, \bibinfo {author} {\bibfnamefont {Y.}~\bibnamefont {Li}}, \bibinfo
  {author} {\bibfnamefont {X.-X.}\ \bibnamefont {Wu}}, \bibinfo {author}
  {\bibfnamefont {H.-B.}\ \bibnamefont {Deng}}, \bibinfo {author}
  {\bibfnamefont {X.}~\bibnamefont {Shi}}, \bibinfo {author} {\bibfnamefont
  {W.-H.}\ \bibnamefont {Fan}}, \bibinfo {author} {\bibfnamefont
  {M.}~\bibnamefont {Li}}, \bibinfo {author} {\bibfnamefont {Y.-B.}\
  \bibnamefont {Huang}}, \bibinfo {author} {\bibfnamefont {T.}~\bibnamefont
  {Qian}}, \bibinfo {author} {\bibfnamefont {P.}~\bibnamefont {Richard}},
  \bibinfo {author} {\bibfnamefont {J.-P.}\ \bibnamefont {Hu}}, \bibinfo
  {author} {\bibfnamefont {S.-H.}\ \bibnamefont {Pan}}, \bibinfo {author}
  {\bibfnamefont {H.-Q.}\ \bibnamefont {Mao}}, \bibinfo {author} {\bibfnamefont
  {Y.-J.}\ \bibnamefont {Sun}}, \ and\ \bibinfo {author} {\bibfnamefont
  {H.}~\bibnamefont {Ding}},\ }\bibfield  {title} {\enquote {\bibinfo {title}
  {{Observation of topological transition in high-${T}_{c}$ superconducting
  monolayer ${\mathrm{FeTe}}_{1\ensuremath{-}x}{\mathrm{Se}}_{x}$ films on
  ${\mathrm{SrTiO}}_{3}(001)$}},}\ }\href {\doibase
  10.1103/PhysRevB.100.155134} {\bibfield  {journal} {\bibinfo  {journal}
  {Phys. Rev. B}\ }\textbf {\bibinfo {volume} {100}},\ \bibinfo {pages}
  {155134} (\bibinfo {year} {2019})}\BibitemShut {NoStop}%
\bibitem [{\citenamefont {Li}\ \emph {et~al.}(2022)\citenamefont {Li},
  \citenamefont {Zhu}, \citenamefont {Fan}, \citenamefont {Cao},\ and\
  \citenamefont {Gao}}]{Li_2022}%
  \BibitemOpen
  \bibfield  {author} {\bibinfo {author} {\bibfnamefont {Geng}\ \bibnamefont
  {Li}}, \bibinfo {author} {\bibfnamefont {Shiyu}\ \bibnamefont {Zhu}},
  \bibinfo {author} {\bibfnamefont {Peng}\ \bibnamefont {Fan}}, \bibinfo
  {author} {\bibfnamefont {Lu}~\bibnamefont {Cao}}, \ and\ \bibinfo {author}
  {\bibfnamefont {Hong-Jun}\ \bibnamefont {Gao}},\ }\bibfield  {title}
  {\enquote {\bibinfo {title} {{Exploring Majorana zero modes in iron-based
  superconductors}},}\ }\href {\doibase 10.1088/1674-1056/ac70c3} {\bibfield
  {journal} {\bibinfo  {journal} {Chinese Physics B}\ }\textbf {\bibinfo
  {volume} {31}},\ \bibinfo {pages} {080301} (\bibinfo {year}
  {2022})}\BibitemShut {NoStop}%
\bibitem [{\citenamefont {Qin}\ \emph {et~al.}(2019)\citenamefont {Qin},
  \citenamefont {Hu}, \citenamefont {Wu}, \citenamefont {Dai}, \citenamefont
  {Fang}, \citenamefont {Zhang},\ and\ \citenamefont {Hu}}]{Qin_2019}%
  \BibitemOpen
  \bibfield  {author} {\bibinfo {author} {\bibfnamefont {Shengshan}\
  \bibnamefont {Qin}}, \bibinfo {author} {\bibfnamefont {Lunhui}\ \bibnamefont
  {Hu}}, \bibinfo {author} {\bibfnamefont {Xianxin}\ \bibnamefont {Wu}},
  \bibinfo {author} {\bibfnamefont {Xia}\ \bibnamefont {Dai}}, \bibinfo
  {author} {\bibfnamefont {Chen}\ \bibnamefont {Fang}}, \bibinfo {author}
  {\bibfnamefont {Fu-Chun}\ \bibnamefont {Zhang}}, \ and\ \bibinfo {author}
  {\bibfnamefont {Jiangping}\ \bibnamefont {Hu}},\ }\bibfield  {title}
  {\enquote {\bibinfo {title} {Topological vortex phase transitions in
  iron-based superconductors},}\ }\href {\doibase 10.1016/j.scib.2019.07.011}
  {\bibfield  {journal} {\bibinfo  {journal} {Science Bulletin}\ }\textbf
  {\bibinfo {volume} {64}},\ \bibinfo {pages} {1207--1214} (\bibinfo {year}
  {2019})}\BibitemShut {NoStop}%
\bibitem [{\citenamefont {Machida}\ and\ \citenamefont
  {Hanaguri}(2023)}]{10.1093/ptep/ptad084}%
  \BibitemOpen
  \bibfield  {author} {\bibinfo {author} {\bibfnamefont {Tadashi}\ \bibnamefont
  {Machida}}\ and\ \bibinfo {author} {\bibfnamefont {Tetsuo}\ \bibnamefont
  {Hanaguri}},\ }\bibfield  {title} {\enquote {\bibinfo {title} {{Searching for
  Majorana quasiparticles at vortex cores in iron-based superconductors}},}\
  }\href {\doibase 10.1093/ptep/ptad084} {\bibfield  {journal} {\bibinfo
  {journal} {Progress of Theoretical and Experimental Physics}\ ,\ \bibinfo
  {pages} {ptad084}} (\bibinfo {year} {2023})}\BibitemShut {NoStop}%
\bibitem [{\citenamefont {Banerjee}\ \emph {et~al.}(2023)\citenamefont
  {Banerjee}, \citenamefont {Lesser}, \citenamefont {Rahman}, \citenamefont
  {Wang}, \citenamefont {Li}, \citenamefont {Kringh\o{}j}, \citenamefont
  {Whiticar}, \citenamefont {Drachmann}, \citenamefont {Thomas}, \citenamefont
  {Wang}, \citenamefont {Manfra}, \citenamefont {Berg}, \citenamefont {Oreg},
  \citenamefont {Stern},\ and\ \citenamefont {Marcus}}]{PhysRevB.107.245304}%
  \BibitemOpen
  \bibfield  {author} {\bibinfo {author} {\bibfnamefont {A.}~\bibnamefont
  {Banerjee}}, \bibinfo {author} {\bibfnamefont {O.}~\bibnamefont {Lesser}},
  \bibinfo {author} {\bibfnamefont {M.~A.}\ \bibnamefont {Rahman}}, \bibinfo
  {author} {\bibfnamefont {H.-R.}\ \bibnamefont {Wang}}, \bibinfo {author}
  {\bibfnamefont {M.-R.}\ \bibnamefont {Li}}, \bibinfo {author} {\bibfnamefont
  {A.}~\bibnamefont {Kringh\o{}j}}, \bibinfo {author} {\bibfnamefont {A.~M.}\
  \bibnamefont {Whiticar}}, \bibinfo {author} {\bibfnamefont {A.~C.~C.}\
  \bibnamefont {Drachmann}}, \bibinfo {author} {\bibfnamefont {C.}~\bibnamefont
  {Thomas}}, \bibinfo {author} {\bibfnamefont {T.}~\bibnamefont {Wang}},
  \bibinfo {author} {\bibfnamefont {M.~J.}\ \bibnamefont {Manfra}}, \bibinfo
  {author} {\bibfnamefont {E.}~\bibnamefont {Berg}}, \bibinfo {author}
  {\bibfnamefont {Y.}~\bibnamefont {Oreg}}, \bibinfo {author} {\bibfnamefont
  {Ady}\ \bibnamefont {Stern}}, \ and\ \bibinfo {author} {\bibfnamefont
  {C.~M.}\ \bibnamefont {Marcus}},\ }\bibfield  {title} {\enquote {\bibinfo
  {title} {{Signatures of a topological phase transition in a planar Josephson
  junction}},}\ }\href {\doibase 10.1103/PhysRevB.107.245304} {\bibfield
  {journal} {\bibinfo  {journal} {Phys. Rev. B}\ }\textbf {\bibinfo {volume}
  {107}},\ \bibinfo {pages} {245304} (\bibinfo {year} {2023})}\BibitemShut
  {NoStop}%
\bibitem [{\citenamefont {Aghaee}\ \emph {et~al.}(2023)\citenamefont {Aghaee}
  \emph {et~al.}}]{PhysRevB.107.245423}%
  \BibitemOpen
  \bibfield  {author} {\bibinfo {author} {\bibfnamefont {Morteza}\ \bibnamefont
  {Aghaee}} \emph {et~al.} (\bibinfo {collaboration} {Microsoft Quantum}),\
  }\bibfield  {title} {\enquote {\bibinfo {title} {{InAs-Al hybrid devices
  passing the topological gap protocol}},}\ }\href {\doibase
  10.1103/PhysRevB.107.245423} {\bibfield  {journal} {\bibinfo  {journal}
  {Phys. Rev. B}\ }\textbf {\bibinfo {volume} {107}},\ \bibinfo {pages}
  {245423} (\bibinfo {year} {2023})}\BibitemShut {NoStop}%
\bibitem [{\citenamefont {{Kitaev}}(2003)}]{2003AnPhy.303....2K}%
  \BibitemOpen
  \bibfield  {author} {\bibinfo {author} {\bibfnamefont {A.~Y.}\ \bibnamefont
  {{Kitaev}}},\ }\bibfield  {title} {\enquote {\bibinfo {title}
  {{Fault-tolerant quantum computation by anyons}},}\ }\href {\doibase
  10.1016/S0003-4916(02)00018-0} {\bibfield  {journal} {\bibinfo  {journal}
  {Annals of Physics}\ }\textbf {\bibinfo {volume} {303}},\ \bibinfo {pages}
  {2--30} (\bibinfo {year} {2003})},\ \Eprint
  {http://arxiv.org/abs/quant-ph/9707021} {quant-ph/9707021} \BibitemShut
  {NoStop}%
\bibitem [{\citenamefont {Nayak}\ \emph {et~al.}(2008)\citenamefont {Nayak},
  \citenamefont {Simon}, \citenamefont {Stern}, \citenamefont {Freedman},\ and\
  \citenamefont {Das~Sarma}}]{RevModPhys.80.1083}%
  \BibitemOpen
  \bibfield  {author} {\bibinfo {author} {\bibfnamefont {Chetan}\ \bibnamefont
  {Nayak}}, \bibinfo {author} {\bibfnamefont {Steven~H.}\ \bibnamefont
  {Simon}}, \bibinfo {author} {\bibfnamefont {Ady}\ \bibnamefont {Stern}},
  \bibinfo {author} {\bibfnamefont {Michael}\ \bibnamefont {Freedman}}, \ and\
  \bibinfo {author} {\bibfnamefont {Sankar}\ \bibnamefont {Das~Sarma}},\
  }\bibfield  {title} {\enquote {\bibinfo {title} {Non-abelian anyons and
  topological quantum computation},}\ }\href {\doibase
  10.1103/RevModPhys.80.1083} {\bibfield  {journal} {\bibinfo  {journal} {Rev.
  Mod. Phys.}\ }\textbf {\bibinfo {volume} {80}},\ \bibinfo {pages}
  {1083--1159} (\bibinfo {year} {2008})}\BibitemShut {NoStop}%
\bibitem [{\citenamefont {Grosfeld}\ \emph {et~al.}(2011)\citenamefont
  {Grosfeld}, \citenamefont {Seradjeh},\ and\ \citenamefont
  {Vishveshwara}}]{Grosfeld_2011}%
  \BibitemOpen
  \bibfield  {author} {\bibinfo {author} {\bibfnamefont {Eytan}\ \bibnamefont
  {Grosfeld}}, \bibinfo {author} {\bibfnamefont {Babak}\ \bibnamefont
  {Seradjeh}}, \ and\ \bibinfo {author} {\bibfnamefont {Smitha}\ \bibnamefont
  {Vishveshwara}},\ }\bibfield  {title} {\enquote {\bibinfo {title} {{Proposed
  Aharonov-Casher interference measurement of non-Abelian vortices in chiral
  p-wave superconductors}},}\ }\href {\doibase 10.1103/physrevb.83.104513}
  {\bibfield  {journal} {\bibinfo  {journal} {Physical Review B}\ }\textbf
  {\bibinfo {volume} {83}} (\bibinfo {year} {2011}),\
  10.1103/physrevb.83.104513}\BibitemShut {NoStop}%
\bibitem [{\citenamefont {Tewari}\ \emph {et~al.}(2007)\citenamefont {Tewari},
  \citenamefont {Das~Sarma}, \citenamefont {Nayak}, \citenamefont {Zhang},\
  and\ \citenamefont {Zoller}}]{PhysRevLett.98.010506}%
  \BibitemOpen
  \bibfield  {author} {\bibinfo {author} {\bibfnamefont {Sumanta}\ \bibnamefont
  {Tewari}}, \bibinfo {author} {\bibfnamefont {S.}~\bibnamefont {Das~Sarma}},
  \bibinfo {author} {\bibfnamefont {Chetan}\ \bibnamefont {Nayak}}, \bibinfo
  {author} {\bibfnamefont {Chuanwei}\ \bibnamefont {Zhang}}, \ and\ \bibinfo
  {author} {\bibfnamefont {P.}~\bibnamefont {Zoller}},\ }\bibfield  {title}
  {\enquote {\bibinfo {title} {{Quantum Computation using Vortices and Majorana
  Zero Modes of a ${p}_{x}+i{p}_{y}$ Superfluid of Fermionic Cold Atoms}},}\
  }\href {\doibase 10.1103/PhysRevLett.98.010506} {\bibfield  {journal}
  {\bibinfo  {journal} {Phys. Rev. Lett.}\ }\textbf {\bibinfo {volume} {98}},\
  \bibinfo {pages} {010506} (\bibinfo {year} {2007})}\BibitemShut {NoStop}%
\bibitem [{\citenamefont {Ma}\ \emph {et~al.}(2020)\citenamefont {Ma},
  \citenamefont {Reichhardt},\ and\ \citenamefont
  {Reichhardt}}]{PhysRevB.101.024514}%
  \BibitemOpen
  \bibfield  {author} {\bibinfo {author} {\bibfnamefont {X.}~\bibnamefont
  {Ma}}, \bibinfo {author} {\bibfnamefont {C.~J.~O.}\ \bibnamefont
  {Reichhardt}}, \ and\ \bibinfo {author} {\bibfnamefont {C.}~\bibnamefont
  {Reichhardt}},\ }\bibfield  {title} {\enquote {\bibinfo {title} {{Braiding
  Majorana fermions and creating quantum logic gates with vortices on a
  periodic pinning structure}},}\ }\href {\doibase 10.1103/PhysRevB.101.024514}
  {\bibfield  {journal} {\bibinfo  {journal} {Phys. Rev. B}\ }\textbf {\bibinfo
  {volume} {101}},\ \bibinfo {pages} {024514} (\bibinfo {year}
  {2020})}\BibitemShut {NoStop}%
\bibitem [{\citenamefont {{Ma}}\ \emph {et~al.}(2021)\citenamefont {{Ma}},
  \citenamefont {{Guan}}, \citenamefont {{Wang}}, \citenamefont {{Li}},
  \citenamefont {{Liu}}, \citenamefont {{Zheng}},\ and\ \citenamefont
  {{Jia}}}]{2021JPhD...54P4003M}%
  \BibitemOpen
  \bibfield  {author} {\bibinfo {author} {\bibfnamefont {Hai-Yang}\
  \bibnamefont {{Ma}}}, \bibinfo {author} {\bibfnamefont {Dandan}\ \bibnamefont
  {{Guan}}}, \bibinfo {author} {\bibfnamefont {Shiyong}\ \bibnamefont
  {{Wang}}}, \bibinfo {author} {\bibfnamefont {Yaoyi}\ \bibnamefont {{Li}}},
  \bibinfo {author} {\bibfnamefont {Canhua}\ \bibnamefont {{Liu}}}, \bibinfo
  {author} {\bibfnamefont {Hao}\ \bibnamefont {{Zheng}}}, \ and\ \bibinfo
  {author} {\bibfnamefont {Jin-Feng}\ \bibnamefont {{Jia}}},\ }\bibfield
  {title} {\enquote {\bibinfo {title} {{Braiding Majorana zero mode in an
  electrically controllable way}},}\ }\href {\doibase 10.1088/1361-6463/ac1371}
  {\bibfield  {journal} {\bibinfo  {journal} {Journal of Physics D Applied
  Physics}\ }\textbf {\bibinfo {volume} {54}},\ \bibinfo {eid} {424003}
  (\bibinfo {year} {2021})}\BibitemShut {NoStop}%
\bibitem [{\citenamefont {Hua}\ \emph {et~al.}(2021)\citenamefont {Hua},
  \citenamefont {Hal\'asz}, \citenamefont {Dumitrescu}, \citenamefont
  {Brahlek},\ and\ \citenamefont {Lawrie}}]{PhysRevB.104.104501}%
  \BibitemOpen
  \bibfield  {author} {\bibinfo {author} {\bibfnamefont {Chengyun}\
  \bibnamefont {Hua}}, \bibinfo {author} {\bibfnamefont {G\'abor~B.}\
  \bibnamefont {Hal\'asz}}, \bibinfo {author} {\bibfnamefont {Eugene}\
  \bibnamefont {Dumitrescu}}, \bibinfo {author} {\bibfnamefont {Matthew}\
  \bibnamefont {Brahlek}}, \ and\ \bibinfo {author} {\bibfnamefont {Benjamin}\
  \bibnamefont {Lawrie}},\ }\bibfield  {title} {\enquote {\bibinfo {title}
  {Optical vortex manipulation for topological quantum computation},}\ }\href
  {\doibase 10.1103/PhysRevB.104.104501} {\bibfield  {journal} {\bibinfo
  {journal} {Phys. Rev. B}\ }\textbf {\bibinfo {volume} {104}},\ \bibinfo
  {pages} {104501} (\bibinfo {year} {2021})}\BibitemShut {NoStop}%
\bibitem [{\citenamefont {Bonderson}\ \emph {et~al.}(2008)\citenamefont
  {Bonderson}, \citenamefont {Freedman},\ and\ \citenamefont
  {Nayak}}]{PhysRevLett.101.010501}%
  \BibitemOpen
  \bibfield  {author} {\bibinfo {author} {\bibfnamefont {Parsa}\ \bibnamefont
  {Bonderson}}, \bibinfo {author} {\bibfnamefont {Michael}\ \bibnamefont
  {Freedman}}, \ and\ \bibinfo {author} {\bibfnamefont {Chetan}\ \bibnamefont
  {Nayak}},\ }\bibfield  {title} {\enquote {\bibinfo {title} {{Measurement-Only
  Topological Quantum Computation}},}\ }\href {\doibase
  10.1103/PhysRevLett.101.010501} {\bibfield  {journal} {\bibinfo  {journal}
  {Phys. Rev. Lett.}\ }\textbf {\bibinfo {volume} {101}},\ \bibinfo {pages}
  {010501} (\bibinfo {year} {2008})}\BibitemShut {NoStop}%
\bibitem [{\citenamefont {Liu}\ \emph {et~al.}(2019)\citenamefont {Liu},
  \citenamefont {Liu}, \citenamefont {Zhang},\ and\ \citenamefont
  {Chiu}}]{PhysRevApplied.12.054035}%
  \BibitemOpen
  \bibfield  {author} {\bibinfo {author} {\bibfnamefont {Chun-Xiao}\
  \bibnamefont {Liu}}, \bibinfo {author} {\bibfnamefont {Dong~E.}\ \bibnamefont
  {Liu}}, \bibinfo {author} {\bibfnamefont {Fu-Chun}\ \bibnamefont {Zhang}}, \
  and\ \bibinfo {author} {\bibfnamefont {Ching-Kai}\ \bibnamefont {Chiu}},\
  }\bibfield  {title} {\enquote {\bibinfo {title} {{Protocol for Reading Out
  Majorana Vortex Qubits and Testing Non-Abelian Statistics}},}\ }\href
  {\doibase 10.1103/PhysRevApplied.12.054035} {\bibfield  {journal} {\bibinfo
  {journal} {Phys. Rev. Appl.}\ }\textbf {\bibinfo {volume} {12}},\ \bibinfo
  {pages} {054035} (\bibinfo {year} {2019})}\BibitemShut {NoStop}%
\bibitem [{\citenamefont {Sbierski}\ \emph {et~al.}(2022)\citenamefont
  {Sbierski}, \citenamefont {Geier}, \citenamefont {Li}, \citenamefont
  {Brahlek}, \citenamefont {Moore},\ and\ \citenamefont
  {Moore}}]{Sbierski_2022}%
  \BibitemOpen
  \bibfield  {author} {\bibinfo {author} {\bibfnamefont {Björn}\ \bibnamefont
  {Sbierski}}, \bibinfo {author} {\bibfnamefont {Max}\ \bibnamefont {Geier}},
  \bibinfo {author} {\bibfnamefont {An-Ping}\ \bibnamefont {Li}}, \bibinfo
  {author} {\bibfnamefont {Matthew}\ \bibnamefont {Brahlek}}, \bibinfo {author}
  {\bibfnamefont {Robert~G.}\ \bibnamefont {Moore}}, \ and\ \bibinfo {author}
  {\bibfnamefont {Joel~E.}\ \bibnamefont {Moore}},\ }\bibfield  {title}
  {\enquote {\bibinfo {title} {{Identifying Majorana vortex modes via nonlocal
  transport}},}\ }\href {\doibase 10.1103/physrevb.106.035413} {\bibfield
  {journal} {\bibinfo  {journal} {Physical Review B}\ }\textbf {\bibinfo
  {volume} {106}} (\bibinfo {year} {2022}),\
  10.1103/physrevb.106.035413}\BibitemShut {NoStop}%
\bibitem [{\citenamefont {Fu}\ and\ \citenamefont
  {Kane}(2009)}]{PhysRevLett.102.216403}%
  \BibitemOpen
  \bibfield  {author} {\bibinfo {author} {\bibfnamefont {Liang}\ \bibnamefont
  {Fu}}\ and\ \bibinfo {author} {\bibfnamefont {C.~L.}\ \bibnamefont {Kane}},\
  }\bibfield  {title} {\enquote {\bibinfo {title} {{Probing Neutral Majorana
  Fermion Edge Modes with Charge Transport}},}\ }\href {\doibase
  10.1103/PhysRevLett.102.216403} {\bibfield  {journal} {\bibinfo  {journal}
  {Phys. Rev. Lett.}\ }\textbf {\bibinfo {volume} {102}},\ \bibinfo {pages}
  {216403} (\bibinfo {year} {2009})}\BibitemShut {NoStop}%
\bibitem [{\citenamefont {Akhmerov}\ \emph {et~al.}(2009)\citenamefont
  {Akhmerov}, \citenamefont {Nilsson},\ and\ \citenamefont
  {Beenakker}}]{PhysRevLett.102.216404}%
  \BibitemOpen
  \bibfield  {author} {\bibinfo {author} {\bibfnamefont {A.~R.}\ \bibnamefont
  {Akhmerov}}, \bibinfo {author} {\bibfnamefont {Johan}\ \bibnamefont
  {Nilsson}}, \ and\ \bibinfo {author} {\bibfnamefont {C.~W.~J.}\ \bibnamefont
  {Beenakker}},\ }\bibfield  {title} {\enquote {\bibinfo {title} {{Electrically
  Detected Interferometry of Majorana Fermions in a Topological Insulator}},}\
  }\href {\doibase 10.1103/PhysRevLett.102.216404} {\bibfield  {journal}
  {\bibinfo  {journal} {Phys. Rev. Lett.}\ }\textbf {\bibinfo {volume} {102}},\
  \bibinfo {pages} {216404} (\bibinfo {year} {2009})}\BibitemShut {NoStop}%
\bibitem [{\citenamefont {Smith}\ \emph {et~al.}(2020)\citenamefont {Smith},
  \citenamefont {Cassidy}, \citenamefont {Reilly}, \citenamefont {Bartlett},\
  and\ \citenamefont {Grimsmo}}]{PRXQuantum.1.020313}%
  \BibitemOpen
  \bibfield  {author} {\bibinfo {author} {\bibfnamefont {Thomas~B.}\
  \bibnamefont {Smith}}, \bibinfo {author} {\bibfnamefont {Maja~C.}\
  \bibnamefont {Cassidy}}, \bibinfo {author} {\bibfnamefont {David~J.}\
  \bibnamefont {Reilly}}, \bibinfo {author} {\bibfnamefont {Stephen~D.}\
  \bibnamefont {Bartlett}}, \ and\ \bibinfo {author} {\bibfnamefont {Arne~L.}\
  \bibnamefont {Grimsmo}},\ }\bibfield  {title} {\enquote {\bibinfo {title}
  {{Dispersive Readout of Majorana Qubits}},}\ }\href {\doibase
  10.1103/PRXQuantum.1.020313} {\bibfield  {journal} {\bibinfo  {journal} {PRX
  Quantum}\ }\textbf {\bibinfo {volume} {1}},\ \bibinfo {pages} {020313}
  (\bibinfo {year} {2020})}\BibitemShut {NoStop}%
\bibitem [{\citenamefont {Razmadze}\ \emph {et~al.}(2019)\citenamefont
  {Razmadze}, \citenamefont {Sabonis}, \citenamefont {Malinowski},
  \citenamefont {M\'enard}, \citenamefont {Pauka}, \citenamefont {Nguyen},
  \citenamefont {van Zanten}, \citenamefont {O\ensuremath{'}Farrell},
  \citenamefont {Suter}, \citenamefont {Krogstrup}, \citenamefont {Kuemmeth},\
  and\ \citenamefont {Marcus}}]{Razmadze2018}%
  \BibitemOpen
  \bibfield  {author} {\bibinfo {author} {\bibfnamefont {Davydas}\ \bibnamefont
  {Razmadze}}, \bibinfo {author} {\bibfnamefont {Deividas}\ \bibnamefont
  {Sabonis}}, \bibinfo {author} {\bibfnamefont {Filip~K.}\ \bibnamefont
  {Malinowski}}, \bibinfo {author} {\bibfnamefont {Gerbold~C.}\ \bibnamefont
  {M\'enard}}, \bibinfo {author} {\bibfnamefont {Sebastian}\ \bibnamefont
  {Pauka}}, \bibinfo {author} {\bibfnamefont {Hung}\ \bibnamefont {Nguyen}},
  \bibinfo {author} {\bibfnamefont {David~M.T.}\ \bibnamefont {van Zanten}},
  \bibinfo {author} {\bibfnamefont {Eoin~C.T.}\ \bibnamefont
  {O\ensuremath{'}Farrell}}, \bibinfo {author} {\bibfnamefont {Judith}\
  \bibnamefont {Suter}}, \bibinfo {author} {\bibfnamefont {Peter}\ \bibnamefont
  {Krogstrup}}, \bibinfo {author} {\bibfnamefont {Ferdinand}\ \bibnamefont
  {Kuemmeth}}, \ and\ \bibinfo {author} {\bibfnamefont {Charles~M.}\
  \bibnamefont {Marcus}},\ }\bibfield  {title} {\enquote {\bibinfo {title}
  {{Radio-Frequency Methods for Majorana-Based Quantum Devices: Fast Charge
  Sensing and Phase-Diagram Mapping}},}\ }\href {\doibase
  10.1103/PhysRevApplied.11.064011} {\bibfield  {journal} {\bibinfo  {journal}
  {Phys. Rev. Appl.}\ }\textbf {\bibinfo {volume} {11}},\ \bibinfo {pages}
  {064011} (\bibinfo {year} {2019})}\BibitemShut {NoStop}%
\bibitem [{\citenamefont {{Bretheau}}\ \emph {et~al.}(2013)\citenamefont
  {{Bretheau}}, \citenamefont {{Girit}}, \citenamefont {{Pothier}},
  \citenamefont {{Esteve}},\ and\ \citenamefont
  {{Urbina}}}]{2013Natur.499..312B}%
  \BibitemOpen
  \bibfield  {author} {\bibinfo {author} {\bibfnamefont {L.}~\bibnamefont
  {{Bretheau}}}, \bibinfo {author} {\bibfnamefont {{\c{C}}.~{\"O}.}\
  \bibnamefont {{Girit}}}, \bibinfo {author} {\bibfnamefont {H.}~\bibnamefont
  {{Pothier}}}, \bibinfo {author} {\bibfnamefont {D.}~\bibnamefont {{Esteve}}},
  \ and\ \bibinfo {author} {\bibfnamefont {C.}~\bibnamefont {{Urbina}}},\
  }\bibfield  {title} {\enquote {\bibinfo {title} {{Exciting Andreev pairs in a
  superconducting atomic contact}},}\ }\href {\doibase 10.1038/nature12315}
  {\bibfield  {journal} {\bibinfo  {journal} {Nature}\ }\textbf {\bibinfo
  {volume} {499}},\ \bibinfo {pages} {312--315} (\bibinfo {year}
  {2013})}\BibitemShut {NoStop}%
\bibitem [{\citenamefont {{van Woerkom}}\ \emph {et~al.}(2017)\citenamefont
  {{van Woerkom}}, \citenamefont {{Proutski}}, \citenamefont {{van Heck}},
  \citenamefont {{Bouman}}, \citenamefont {{V{\"a}yrynen}}, \citenamefont
  {{Glazman}}, \citenamefont {{Krogstrup}}, \citenamefont {{Nyg{\r{a}}rd}},
  \citenamefont {{Kouwenhoven}},\ and\ \citenamefont
  {{Geresdi}}}]{Woerkom2016}%
  \BibitemOpen
  \bibfield  {author} {\bibinfo {author} {\bibfnamefont {David~J.}\
  \bibnamefont {{van Woerkom}}}, \bibinfo {author} {\bibfnamefont {Alex}\
  \bibnamefont {{Proutski}}}, \bibinfo {author} {\bibfnamefont {Bernard}\
  \bibnamefont {{van Heck}}}, \bibinfo {author} {\bibfnamefont {Dani{\"e}l}\
  \bibnamefont {{Bouman}}}, \bibinfo {author} {\bibfnamefont {Jukka~I.}\
  \bibnamefont {{V{\"a}yrynen}}}, \bibinfo {author} {\bibfnamefont {Leonid~I.}\
  \bibnamefont {{Glazman}}}, \bibinfo {author} {\bibfnamefont {Peter}\
  \bibnamefont {{Krogstrup}}}, \bibinfo {author} {\bibfnamefont {Jesper}\
  \bibnamefont {{Nyg{\r{a}}rd}}}, \bibinfo {author} {\bibfnamefont {Leo~P.}\
  \bibnamefont {{Kouwenhoven}}}, \ and\ \bibinfo {author} {\bibfnamefont
  {Attila}\ \bibnamefont {{Geresdi}}},\ }\bibfield  {title} {\enquote {\bibinfo
  {title} {{Microwave spectroscopy of spinful Andreev bound states in ballistic
  semiconductor Josephson junctions}},}\ }\href {\doibase 10.1038/nphys4150}
  {\bibfield  {journal} {\bibinfo  {journal} {Nat. Phys.}\ }\textbf {\bibinfo
  {volume} {13}},\ \bibinfo {pages} {876--881} (\bibinfo {year}
  {2017})}\BibitemShut {NoStop}%
\bibitem [{\citenamefont {{Tosi}}\ \emph {et~al.}(2019)\citenamefont {{Tosi}},
  \citenamefont {{Metzger}}, \citenamefont {{Goffman}}, \citenamefont
  {{Urbina}}, \citenamefont {{Pothier}}, \citenamefont {{Park}}, \citenamefont
  {{Yeyati}}, \citenamefont {{Nyg{\^a}rd}},\ and\ \citenamefont
  {{Krogstrup}}}]{2019PhRvX...9a1010T}%
  \BibitemOpen
  \bibfield  {author} {\bibinfo {author} {\bibfnamefont {L.}~\bibnamefont
  {{Tosi}}}, \bibinfo {author} {\bibfnamefont {C.}~\bibnamefont {{Metzger}}},
  \bibinfo {author} {\bibfnamefont {M.~F.}\ \bibnamefont {{Goffman}}}, \bibinfo
  {author} {\bibfnamefont {C.}~\bibnamefont {{Urbina}}}, \bibinfo {author}
  {\bibfnamefont {H.}~\bibnamefont {{Pothier}}}, \bibinfo {author}
  {\bibfnamefont {Sunghun}\ \bibnamefont {{Park}}}, \bibinfo {author}
  {\bibfnamefont {A.~Levy}\ \bibnamefont {{Yeyati}}}, \bibinfo {author}
  {\bibfnamefont {J.}~\bibnamefont {{Nyg{\^a}rd}}}, \ and\ \bibinfo {author}
  {\bibfnamefont {P.}~\bibnamefont {{Krogstrup}}},\ }\bibfield  {title}
  {\enquote {\bibinfo {title} {{Spin-Orbit Splitting of Andreev States Revealed
  by Microwave Spectroscopy}},}\ }\href {\doibase 10.1103/PhysRevX.9.011010}
  {\bibfield  {journal} {\bibinfo  {journal} {Phys. Rev. X}\ }\textbf {\bibinfo
  {volume} {9}},\ \bibinfo {eid} {011010} (\bibinfo {year} {2019})}\BibitemShut
  {NoStop}%
\bibitem [{\citenamefont {Hays}\ \emph {et~al.}(2020)\citenamefont {Hays},
  \citenamefont {Fatemi}, \citenamefont {Serniak}, \citenamefont {Bouman},
  \citenamefont {Diamond}, \citenamefont {de~Lange}, \citenamefont {Krogstrup},
  \citenamefont {Nyg{\aa}rd}, \citenamefont {Geresdi},\ and\ \citenamefont
  {Devoret}}]{hays2020continuous}%
  \BibitemOpen
  \bibfield  {author} {\bibinfo {author} {\bibfnamefont {M}~\bibnamefont
  {Hays}}, \bibinfo {author} {\bibfnamefont {V}~\bibnamefont {Fatemi}},
  \bibinfo {author} {\bibfnamefont {K}~\bibnamefont {Serniak}}, \bibinfo
  {author} {\bibfnamefont {D}~\bibnamefont {Bouman}}, \bibinfo {author}
  {\bibfnamefont {S}~\bibnamefont {Diamond}}, \bibinfo {author} {\bibfnamefont
  {G}~\bibnamefont {de~Lange}}, \bibinfo {author} {\bibfnamefont
  {P}~\bibnamefont {Krogstrup}}, \bibinfo {author} {\bibfnamefont
  {J}~\bibnamefont {Nyg{\aa}rd}}, \bibinfo {author} {\bibfnamefont
  {A}~\bibnamefont {Geresdi}}, \ and\ \bibinfo {author} {\bibfnamefont
  {MH}~\bibnamefont {Devoret}},\ }\bibfield  {title} {\enquote {\bibinfo
  {title} {Continuous monitoring of a trapped superconducting spin},}\ }\href
  {\doibase https://doi.org/10.1038/s41567-020-0952-3} {\bibfield  {journal}
  {\bibinfo  {journal} {Nature Physics}\ }\textbf {\bibinfo {volume} {16}},\
  \bibinfo {pages} {1103--1107} (\bibinfo {year} {2020})}\BibitemShut {NoStop}%
\bibitem [{\citenamefont {Hays}\ \emph {et~al.}(2021)\citenamefont {Hays},
  \citenamefont {Fatemi}, \citenamefont {Bouman}, \citenamefont {Cerrillo},
  \citenamefont {Diamond}, \citenamefont {Serniak}, \citenamefont {Connolly},
  \citenamefont {Krogstrup}, \citenamefont {Nyg{\aa}rd}, \citenamefont
  {Levy~Yeyati} \emph {et~al.}}]{hays2021coherent}%
  \BibitemOpen
  \bibfield  {author} {\bibinfo {author} {\bibfnamefont {M}~\bibnamefont
  {Hays}}, \bibinfo {author} {\bibfnamefont {V}~\bibnamefont {Fatemi}},
  \bibinfo {author} {\bibfnamefont {D}~\bibnamefont {Bouman}}, \bibinfo
  {author} {\bibfnamefont {J}~\bibnamefont {Cerrillo}}, \bibinfo {author}
  {\bibfnamefont {S}~\bibnamefont {Diamond}}, \bibinfo {author} {\bibfnamefont
  {K}~\bibnamefont {Serniak}}, \bibinfo {author} {\bibfnamefont
  {T}~\bibnamefont {Connolly}}, \bibinfo {author} {\bibfnamefont
  {P}~\bibnamefont {Krogstrup}}, \bibinfo {author} {\bibfnamefont
  {J}~\bibnamefont {Nyg{\aa}rd}}, \bibinfo {author} {\bibfnamefont
  {A}~\bibnamefont {Levy~Yeyati}},  \emph {et~al.},\ }\bibfield  {title}
  {\enquote {\bibinfo {title} {{Coherent manipulation of an Andreev spin
  qubit}},}\ }\href {\doibase https://doi.org/10.1126/science.abf0345}
  {\bibfield  {journal} {\bibinfo  {journal} {Science}\ }\textbf {\bibinfo
  {volume} {373}},\ \bibinfo {pages} {430--433} (\bibinfo {year}
  {2021})}\BibitemShut {NoStop}%
\bibitem [{\citenamefont {Fatemi}\ \emph {et~al.}(2022)\citenamefont {Fatemi},
  \citenamefont {Kurilovich}, \citenamefont {Hays}, \citenamefont {Bouman},
  \citenamefont {Connolly}, \citenamefont {Diamond}, \citenamefont {Frattini},
  \citenamefont {Kurilovich}, \citenamefont {Krogstrup}, \citenamefont {Nygard}
  \emph {et~al.}}]{fatemi2021microwave}%
  \BibitemOpen
  \bibfield  {author} {\bibinfo {author} {\bibfnamefont {V}~\bibnamefont
  {Fatemi}}, \bibinfo {author} {\bibfnamefont {PD}~\bibnamefont {Kurilovich}},
  \bibinfo {author} {\bibfnamefont {M}~\bibnamefont {Hays}}, \bibinfo {author}
  {\bibfnamefont {D}~\bibnamefont {Bouman}}, \bibinfo {author} {\bibfnamefont
  {T}~\bibnamefont {Connolly}}, \bibinfo {author} {\bibfnamefont
  {S}~\bibnamefont {Diamond}}, \bibinfo {author} {\bibfnamefont
  {NE}~\bibnamefont {Frattini}}, \bibinfo {author} {\bibfnamefont
  {VD}~\bibnamefont {Kurilovich}}, \bibinfo {author} {\bibfnamefont
  {P}~\bibnamefont {Krogstrup}}, \bibinfo {author} {\bibfnamefont
  {J}~\bibnamefont {Nygard}},  \emph {et~al.},\ }\bibfield  {title} {\enquote
  {\bibinfo {title} {{Microwave Susceptibility Observation of Interacting
  Many-Body Andreev States}},}\ }\href {\doibase
  10.1103/PhysRevLett.129.227701} {\bibfield  {journal} {\bibinfo  {journal}
  {Phys. Rev. Lett.}\ }\textbf {\bibinfo {volume} {129}},\ \bibinfo {pages}
  {227701} (\bibinfo {year} {2022})}\BibitemShut {NoStop}%
\bibitem [{\citenamefont {Matute-Ca\~nadas}\ \emph {et~al.}(2022)\citenamefont
  {Matute-Ca\~nadas}, \citenamefont {Metzger}, \citenamefont {Park},
  \citenamefont {Tosi}, \citenamefont {Krogstrup}, \citenamefont {Nyg\aa{}rd},
  \citenamefont {Goffman}, \citenamefont {Urbina}, \citenamefont {Pothier},\
  and\ \citenamefont {Yeyati}}]{PhysRevLett.128.197702}%
  \BibitemOpen
  \bibfield  {author} {\bibinfo {author} {\bibfnamefont {F.~J.}\ \bibnamefont
  {Matute-Ca\~nadas}}, \bibinfo {author} {\bibfnamefont {C.}~\bibnamefont
  {Metzger}}, \bibinfo {author} {\bibfnamefont {Sunghun}\ \bibnamefont {Park}},
  \bibinfo {author} {\bibfnamefont {L.}~\bibnamefont {Tosi}}, \bibinfo {author}
  {\bibfnamefont {P.}~\bibnamefont {Krogstrup}}, \bibinfo {author}
  {\bibfnamefont {J.}~\bibnamefont {Nyg\aa{}rd}}, \bibinfo {author}
  {\bibfnamefont {M.~F.}\ \bibnamefont {Goffman}}, \bibinfo {author}
  {\bibfnamefont {C.}~\bibnamefont {Urbina}}, \bibinfo {author} {\bibfnamefont
  {H.}~\bibnamefont {Pothier}}, \ and\ \bibinfo {author} {\bibfnamefont
  {A.~Levy}\ \bibnamefont {Yeyati}},\ }\bibfield  {title} {\enquote {\bibinfo
  {title} {{Signatures of Interactions in the Andreev Spectrum of Nanowire
  Josephson Junctions}},}\ }\href {\doibase 10.1103/PhysRevLett.128.197702}
  {\bibfield  {journal} {\bibinfo  {journal} {Phys. Rev. Lett.}\ }\textbf
  {\bibinfo {volume} {128}},\ \bibinfo {pages} {197702} (\bibinfo {year}
  {2022})}\BibitemShut {NoStop}%
\bibitem [{\citenamefont {Wesdorp}\ \emph {et~al.}(2022)\citenamefont
  {Wesdorp}, \citenamefont {Matute-Caňadas}, \citenamefont {Vaartjes},
  \citenamefont {Grünhaupt}, \citenamefont {Laeven}, \citenamefont {Roelofs},
  \citenamefont {Splitthoff}, \citenamefont {Pita-Vidal}, \citenamefont
  {Bargerbos}, \citenamefont {van Woerkom}, \citenamefont {Krogstrup},
  \citenamefont {Kouwenhoven}, \citenamefont {Andersen}, \citenamefont
  {Yeyati}, \citenamefont {van Heck},\ and\ \citenamefont
  {de~Lange}}]{arxiv.2208.11198}%
  \BibitemOpen
  \bibfield  {author} {\bibinfo {author} {\bibfnamefont {J.~J.}\ \bibnamefont
  {Wesdorp}}, \bibinfo {author} {\bibfnamefont {F.~J.}\ \bibnamefont
  {Matute-Caňadas}}, \bibinfo {author} {\bibfnamefont {A.}~\bibnamefont
  {Vaartjes}}, \bibinfo {author} {\bibfnamefont {L.}~\bibnamefont
  {Grünhaupt}}, \bibinfo {author} {\bibfnamefont {T.}~\bibnamefont {Laeven}},
  \bibinfo {author} {\bibfnamefont {S.}~\bibnamefont {Roelofs}}, \bibinfo
  {author} {\bibfnamefont {L.~J.}\ \bibnamefont {Splitthoff}}, \bibinfo
  {author} {\bibfnamefont {M.}~\bibnamefont {Pita-Vidal}}, \bibinfo {author}
  {\bibfnamefont {A.}~\bibnamefont {Bargerbos}}, \bibinfo {author}
  {\bibfnamefont {D.~J.}\ \bibnamefont {van Woerkom}}, \bibinfo {author}
  {\bibfnamefont {P.}~\bibnamefont {Krogstrup}}, \bibinfo {author}
  {\bibfnamefont {L.~P.}\ \bibnamefont {Kouwenhoven}}, \bibinfo {author}
  {\bibfnamefont {C.~K.}\ \bibnamefont {Andersen}}, \bibinfo {author}
  {\bibfnamefont {A.~Levy}\ \bibnamefont {Yeyati}}, \bibinfo {author}
  {\bibfnamefont {B.}~\bibnamefont {van Heck}}, \ and\ \bibinfo {author}
  {\bibfnamefont {G.}~\bibnamefont {de~Lange}},\ }\bibfield  {title} {\enquote
  {\bibinfo {title} {{Microwave spectroscopy of interacting Andreev spins}},}\
  }\href {\doibase 10.48550/ARXIV.2208.11198} {\  (\bibinfo {year} {2022}),\
  10.48550/ARXIV.2208.11198}\BibitemShut {NoStop}%
\bibitem [{\citenamefont {Yavilberg}\ \emph {et~al.}(2015)\citenamefont
  {Yavilberg}, \citenamefont {Ginossar},\ and\ \citenamefont
  {Grosfeld}}]{Yavilberg2015}%
  \BibitemOpen
  \bibfield  {author} {\bibinfo {author} {\bibfnamefont {Konstantin}\
  \bibnamefont {Yavilberg}}, \bibinfo {author} {\bibfnamefont {Eran}\
  \bibnamefont {Ginossar}}, \ and\ \bibinfo {author} {\bibfnamefont {Eytan}\
  \bibnamefont {Grosfeld}},\ }\bibfield  {title} {\enquote {\bibinfo {title}
  {{Fermion parity measurement and control in Majorana circuit quantum
  electrodynamics}},}\ }\href {\doibase 10.1103/PhysRevB.92.075143} {\bibfield
  {journal} {\bibinfo  {journal} {Phys. Rev. B}\ }\textbf {\bibinfo {volume}
  {92}},\ \bibinfo {pages} {075143} (\bibinfo {year} {2015})}\BibitemShut
  {NoStop}%
\bibitem [{\citenamefont {{Ginossar}}\ and\ \citenamefont
  {{Grosfeld}}(2014)}]{2014NatCo...5E4772G}%
  \BibitemOpen
  \bibfield  {author} {\bibinfo {author} {\bibfnamefont {E.}~\bibnamefont
  {{Ginossar}}}\ and\ \bibinfo {author} {\bibfnamefont {E.}~\bibnamefont
  {{Grosfeld}}},\ }\bibfield  {title} {\enquote {\bibinfo {title} {{Microwave
  transitions as a signature of coherent parity mixing effects in the
  Majorana-transmon qubit}},}\ }\href {\doibase 10.1038/ncomms5772} {\bibfield
  {journal} {\bibinfo  {journal} {Nat. Commun.}\ }\textbf {\bibinfo {volume}
  {5}},\ \bibinfo {eid} {4772} (\bibinfo {year} {2014})},\ \Eprint
  {http://arxiv.org/abs/1307.1159} {arXiv:1307.1159 [cond-mat.mes-hall]}
  \BibitemShut {NoStop}%
\bibitem [{\citenamefont {Dmytruk}\ \emph {et~al.}(2015)\citenamefont
  {Dmytruk}, \citenamefont {Trif},\ and\ \citenamefont
  {Simon}}]{PhysRevB.92.245432}%
  \BibitemOpen
  \bibfield  {author} {\bibinfo {author} {\bibfnamefont {Olesia}\ \bibnamefont
  {Dmytruk}}, \bibinfo {author} {\bibfnamefont {Mircea}\ \bibnamefont {Trif}},
  \ and\ \bibinfo {author} {\bibfnamefont {Pascal}\ \bibnamefont {Simon}},\
  }\bibfield  {title} {\enquote {\bibinfo {title} {Cavity quantum
  electrodynamics with mesoscopic topological superconductors},}\ }\href
  {\doibase 10.1103/PhysRevB.92.245432} {\bibfield  {journal} {\bibinfo
  {journal} {Phys. Rev. B}\ }\textbf {\bibinfo {volume} {92}},\ \bibinfo
  {pages} {245432} (\bibinfo {year} {2015})}\BibitemShut {NoStop}%
\bibitem [{\citenamefont {V{\"{a}}yrynen}\ \emph {et~al.}(2015)\citenamefont
  {V{\"{a}}yrynen}, \citenamefont {Rastelli}, \citenamefont {Belzig},\ and\
  \citenamefont {Glazman}}]{Vayrynen2015}%
  \BibitemOpen
  \bibfield  {author} {\bibinfo {author} {\bibfnamefont {Jukka~I.}\
  \bibnamefont {V{\"{a}}yrynen}}, \bibinfo {author} {\bibfnamefont {Gianluca}\
  \bibnamefont {Rastelli}}, \bibinfo {author} {\bibfnamefont {Wolfgang}\
  \bibnamefont {Belzig}}, \ and\ \bibinfo {author} {\bibfnamefont {Leonid~I.}\
  \bibnamefont {Glazman}},\ }\bibfield  {title} {\enquote {\bibinfo {title}
  {{Microwave signatures of Majorana states in a topological Josephson
  junction}},}\ }\href {\doibase 10.1103/PhysRevB.92.134508} {\bibfield
  {journal} {\bibinfo  {journal} {Phys. Rev. B}\ }\textbf {\bibinfo {volume}
  {92}},\ \bibinfo {pages} {134508} (\bibinfo {year} {2015})}\BibitemShut
  {NoStop}%
\bibitem [{\citenamefont {Blais}\ \emph {et~al.}(2021)\citenamefont {Blais},
  \citenamefont {Grimsmo}, \citenamefont {Girvin},\ and\ \citenamefont
  {Wallraff}}]{Blais2020}%
  \BibitemOpen
  \bibfield  {author} {\bibinfo {author} {\bibfnamefont {Alexandre}\
  \bibnamefont {Blais}}, \bibinfo {author} {\bibfnamefont {Arne~L.}\
  \bibnamefont {Grimsmo}}, \bibinfo {author} {\bibfnamefont {S.~M.}\
  \bibnamefont {Girvin}}, \ and\ \bibinfo {author} {\bibfnamefont {Andreas}\
  \bibnamefont {Wallraff}},\ }\bibfield  {title} {\enquote {\bibinfo {title}
  {{Circuit quantum electrodynamics}},}\ }\href {\doibase
  10.1103/RevModPhys.93.025005} {\bibfield  {journal} {\bibinfo  {journal}
  {Rev. Mod. Phys.}\ }\textbf {\bibinfo {volume} {93}},\ \bibinfo {pages}
  {025005} (\bibinfo {year} {2021})}\BibitemShut {NoStop}%
\bibitem [{\citenamefont {Göppl}\ \emph {et~al.}(2008)\citenamefont {Göppl},
  \citenamefont {Fragner}, \citenamefont {Baur}, \citenamefont {Bianchetti},
  \citenamefont {Filipp}, \citenamefont {Fink}, \citenamefont {Leek},
  \citenamefont {Puebla}, \citenamefont {Steffen},\ and\ \citenamefont
  {Wallraff}}]{10.1063/1.3010859}%
  \BibitemOpen
  \bibfield  {author} {\bibinfo {author} {\bibfnamefont {M.}~\bibnamefont
  {Göppl}}, \bibinfo {author} {\bibfnamefont {A.}~\bibnamefont {Fragner}},
  \bibinfo {author} {\bibfnamefont {M.}~\bibnamefont {Baur}}, \bibinfo {author}
  {\bibfnamefont {R.}~\bibnamefont {Bianchetti}}, \bibinfo {author}
  {\bibfnamefont {S.}~\bibnamefont {Filipp}}, \bibinfo {author} {\bibfnamefont
  {J.~M.}\ \bibnamefont {Fink}}, \bibinfo {author} {\bibfnamefont {P.~J.}\
  \bibnamefont {Leek}}, \bibinfo {author} {\bibfnamefont {G.}~\bibnamefont
  {Puebla}}, \bibinfo {author} {\bibfnamefont {L.}~\bibnamefont {Steffen}}, \
  and\ \bibinfo {author} {\bibfnamefont {A.}~\bibnamefont {Wallraff}},\
  }\bibfield  {title} {\enquote {\bibinfo {title} {{Coplanar waveguide
  resonators for circuit quantum electrodynamics}},}\ }\href {\doibase
  10.1063/1.3010859} {\bibfield  {journal} {\bibinfo  {journal} {Journal of
  Applied Physics}\ }\textbf {\bibinfo {volume} {104}} (\bibinfo {year}
  {2008}),\ 10.1063/1.3010859},\ \bibinfo {note} {113904}\BibitemShut {NoStop}%
\bibitem [{Sup()}]{SuppMat}%
  \BibitemOpen
  \href@noop {} {}\bibinfo {note} {See supplementary materials for
  details.}\BibitemShut {Stop}%
\bibitem [{\citenamefont {Zhang}\ \emph
  {et~al.}(2019{\natexlab{b}})\citenamefont {Zhang}, \citenamefont {Cole},\
  and\ \citenamefont {Das~Sarma}}]{PhysRevLett.122.187001}%
  \BibitemOpen
  \bibfield  {author} {\bibinfo {author} {\bibfnamefont {Rui-Xing}\
  \bibnamefont {Zhang}}, \bibinfo {author} {\bibfnamefont {William~S.}\
  \bibnamefont {Cole}}, \ and\ \bibinfo {author} {\bibfnamefont
  {S.}~\bibnamefont {Das~Sarma}},\ }\bibfield  {title} {\enquote {\bibinfo
  {title} {{Helical Hinge Majorana Modes in Iron-Based Superconductors}},}\
  }\href {\doibase 10.1103/PhysRevLett.122.187001} {\bibfield  {journal}
  {\bibinfo  {journal} {Phys. Rev. Lett.}\ }\textbf {\bibinfo {volume} {122}},\
  \bibinfo {pages} {187001} (\bibinfo {year} {2019}{\natexlab{b}})}\BibitemShut
  {NoStop}%
\bibitem [{\citenamefont {Ghazaryan}\ \emph {et~al.}(2020)\citenamefont
  {Ghazaryan}, \citenamefont {Lopes}, \citenamefont {Hosur}, \citenamefont
  {Gilbert},\ and\ \citenamefont {Ghaemi}}]{PhysRevB.101.020504}%
  \BibitemOpen
  \bibfield  {author} {\bibinfo {author} {\bibfnamefont {Areg}\ \bibnamefont
  {Ghazaryan}}, \bibinfo {author} {\bibfnamefont {P.~L.~S.}\ \bibnamefont
  {Lopes}}, \bibinfo {author} {\bibfnamefont {Pavan}\ \bibnamefont {Hosur}},
  \bibinfo {author} {\bibfnamefont {Matthew~J.}\ \bibnamefont {Gilbert}}, \
  and\ \bibinfo {author} {\bibfnamefont {Pouyan}\ \bibnamefont {Ghaemi}},\
  }\bibfield  {title} {\enquote {\bibinfo {title} {{Effect of Zeeman coupling
  on the Majorana vortex modes in iron-based topological superconductors}},}\
  }\href {\doibase 10.1103/PhysRevB.101.020504} {\bibfield  {journal} {\bibinfo
   {journal} {Phys. Rev. B}\ }\textbf {\bibinfo {volume} {101}},\ \bibinfo
  {pages} {020504} (\bibinfo {year} {2020})}\BibitemShut {NoStop}%
\bibitem [{\citenamefont {Chiu}\ \emph {et~al.}(2020)\citenamefont {Chiu},
  \citenamefont {Machida}, \citenamefont {Huang}, \citenamefont {Hanaguri},\
  and\ \citenamefont {Zhang}}]{doi:10.1126/sciadv.aay0443}%
  \BibitemOpen
  \bibfield  {author} {\bibinfo {author} {\bibfnamefont {Ching-Kai}\
  \bibnamefont {Chiu}}, \bibinfo {author} {\bibfnamefont {T.}~\bibnamefont
  {Machida}}, \bibinfo {author} {\bibfnamefont {Yingyi}\ \bibnamefont {Huang}},
  \bibinfo {author} {\bibfnamefont {T.}~\bibnamefont {Hanaguri}}, \ and\
  \bibinfo {author} {\bibfnamefont {Fu-Chun}\ \bibnamefont {Zhang}},\
  }\bibfield  {title} {\enquote {\bibinfo {title} {{Scalable Majorana vortex
  modes in iron-based superconductors}},}\ }\href {\doibase
  10.1126/sciadv.aay0443} {\bibfield  {journal} {\bibinfo  {journal} {Science
  Advances}\ }\textbf {\bibinfo {volume} {6}},\ \bibinfo {pages} {eaay0443}
  (\bibinfo {year} {2020})}\BibitemShut {NoStop}%
\bibitem [{\citenamefont {Hou}\ and\ \citenamefont
  {Klinovaja}(2021)}]{HouKlinovaja2021}%
  \BibitemOpen
  \bibfield  {author} {\bibinfo {author} {\bibfnamefont {Zhe}\ \bibnamefont
  {Hou}}\ and\ \bibinfo {author} {\bibfnamefont {Jelena}\ \bibnamefont
  {Klinovaja}},\ }\href {\doibase 10.48550/ARXIV.2109.08200} {\enquote
  {\bibinfo {title} {{Zero-energy Andreev bound states in iron-based
  superconductor Fe(Te,Se)}},}\ } (\bibinfo {year} {2021})\BibitemShut
  {NoStop}%
\bibitem [{\citenamefont {Barik}\ and\ \citenamefont
  {Sau}(2022)}]{PhysRevB.105.035128}%
  \BibitemOpen
  \bibfield  {author} {\bibinfo {author} {\bibfnamefont {Tamoghna}\
  \bibnamefont {Barik}}\ and\ \bibinfo {author} {\bibfnamefont {Jay~D.}\
  \bibnamefont {Sau}},\ }\bibfield  {title} {\enquote {\bibinfo {title}
  {Signatures of nontopological patches on the surface of topological
  insulators},}\ }\href {\doibase 10.1103/PhysRevB.105.035128} {\bibfield
  {journal} {\bibinfo  {journal} {Phys. Rev. B}\ }\textbf {\bibinfo {volume}
  {105}},\ \bibinfo {pages} {035128} (\bibinfo {year} {2022})}\BibitemShut
  {NoStop}%
\bibitem [{\citenamefont {Blatter}\ \emph {et~al.}(1996)\citenamefont
  {Blatter}, \citenamefont {Feigel'man}, \citenamefont {Geshkenbein},
  \citenamefont {Larkin},\ and\ \citenamefont {van
  Otterlo}}]{PhysRevLett.77.566}%
  \BibitemOpen
  \bibfield  {author} {\bibinfo {author} {\bibfnamefont {Gianni}\ \bibnamefont
  {Blatter}}, \bibinfo {author} {\bibfnamefont {Mikhail}\ \bibnamefont
  {Feigel'man}}, \bibinfo {author} {\bibfnamefont {Vadim}\ \bibnamefont
  {Geshkenbein}}, \bibinfo {author} {\bibfnamefont {Anatoli}\ \bibnamefont
  {Larkin}}, \ and\ \bibinfo {author} {\bibfnamefont {Anne}\ \bibnamefont {van
  Otterlo}},\ }\bibfield  {title} {\enquote {\bibinfo {title} {{Electrostatics
  of Vortices in Type-II Superconductors}},}\ }\href {\doibase
  10.1103/PhysRevLett.77.566} {\bibfield  {journal} {\bibinfo  {journal} {Phys.
  Rev. Lett.}\ }\textbf {\bibinfo {volume} {77}},\ \bibinfo {pages} {566--569}
  (\bibinfo {year} {1996})}\BibitemShut {NoStop}%
\bibitem [{\citenamefont {Groth}\ \emph {et~al.}(2014)\citenamefont {Groth},
  \citenamefont {Wimmer}, \citenamefont {Akhmerov},\ and\ \citenamefont
  {Waintal}}]{groth2014kwant}%
  \BibitemOpen
  \bibfield  {author} {\bibinfo {author} {\bibfnamefont {Christoph~W}\
  \bibnamefont {Groth}}, \bibinfo {author} {\bibfnamefont {Michael}\
  \bibnamefont {Wimmer}}, \bibinfo {author} {\bibfnamefont {Anton~R}\
  \bibnamefont {Akhmerov}}, \ and\ \bibinfo {author} {\bibfnamefont {Xavier}\
  \bibnamefont {Waintal}},\ }\bibfield  {title} {\enquote {\bibinfo {title}
  {Kwant: a software package for quantum transport},}\ }\href@noop {}
  {\bibfield  {journal} {\bibinfo  {journal} {New Journal of Physics}\ }\textbf
  {\bibinfo {volume} {16}},\ \bibinfo {pages} {063065} (\bibinfo {year}
  {2014})}\BibitemShut {NoStop}%
\bibitem [{\citenamefont {Noguchi}\ \emph {et~al.}(2019)\citenamefont
  {Noguchi}, \citenamefont {Dominjon}, \citenamefont {Kroug}, \citenamefont
  {Mima},\ and\ \citenamefont {Otani}}]{8665969}%
  \BibitemOpen
  \bibfield  {author} {\bibinfo {author} {\bibfnamefont {Takashi}\ \bibnamefont
  {Noguchi}}, \bibinfo {author} {\bibfnamefont {Agnes}\ \bibnamefont
  {Dominjon}}, \bibinfo {author} {\bibfnamefont {Matthias}\ \bibnamefont
  {Kroug}}, \bibinfo {author} {\bibfnamefont {Satoru}\ \bibnamefont {Mima}}, \
  and\ \bibinfo {author} {\bibfnamefont {Chiko}\ \bibnamefont {Otani}},\
  }\bibfield  {title} {\enquote {\bibinfo {title} {{Characteristics of Very
  High Q Nb Superconducting Resonators for Microwave Kinetic Inductance
  Detectors}},}\ }\href {\doibase 10.1109/TASC.2019.2904592} {\bibfield
  {journal} {\bibinfo  {journal} {IEEE Transactions on Applied
  Superconductivity}\ }\textbf {\bibinfo {volume} {29}},\ \bibinfo {pages}
  {1--5} (\bibinfo {year} {2019})}\BibitemShut {NoStop}%
\bibitem [{\citenamefont {Vool}\ \emph {et~al.}(2014)\citenamefont {Vool},
  \citenamefont {Pop}, \citenamefont {Sliwa}, \citenamefont {Abdo},
  \citenamefont {Wang}, \citenamefont {Brecht}, \citenamefont {Gao},
  \citenamefont {Shankar}, \citenamefont {Hatridge}, \citenamefont {Catelani},
  \citenamefont {Mirrahimi}, \citenamefont {Frunzio}, \citenamefont
  {Schoelkopf}, \citenamefont {Glazman},\ and\ \citenamefont
  {Devoret}}]{PhysRevLett.113.247001}%
  \BibitemOpen
  \bibfield  {author} {\bibinfo {author} {\bibfnamefont {U.}~\bibnamefont
  {Vool}}, \bibinfo {author} {\bibfnamefont {I.~M.}\ \bibnamefont {Pop}},
  \bibinfo {author} {\bibfnamefont {K.}~\bibnamefont {Sliwa}}, \bibinfo
  {author} {\bibfnamefont {B.}~\bibnamefont {Abdo}}, \bibinfo {author}
  {\bibfnamefont {C.}~\bibnamefont {Wang}}, \bibinfo {author} {\bibfnamefont
  {T.}~\bibnamefont {Brecht}}, \bibinfo {author} {\bibfnamefont {Y.~Y.}\
  \bibnamefont {Gao}}, \bibinfo {author} {\bibfnamefont {S.}~\bibnamefont
  {Shankar}}, \bibinfo {author} {\bibfnamefont {M.}~\bibnamefont {Hatridge}},
  \bibinfo {author} {\bibfnamefont {G.}~\bibnamefont {Catelani}}, \bibinfo
  {author} {\bibfnamefont {M.}~\bibnamefont {Mirrahimi}}, \bibinfo {author}
  {\bibfnamefont {L.}~\bibnamefont {Frunzio}}, \bibinfo {author} {\bibfnamefont
  {R.~J.}\ \bibnamefont {Schoelkopf}}, \bibinfo {author} {\bibfnamefont
  {L.~I.}\ \bibnamefont {Glazman}}, \ and\ \bibinfo {author} {\bibfnamefont
  {M.~H.}\ \bibnamefont {Devoret}},\ }\bibfield  {title} {\enquote {\bibinfo
  {title} {{Non-Poissonian Quantum Jumps of a Fluxonium Qubit due to
  Quasiparticle Excitations}},}\ }\href {\doibase
  10.1103/PhysRevLett.113.247001} {\bibfield  {journal} {\bibinfo  {journal}
  {Phys. Rev. Lett.}\ }\textbf {\bibinfo {volume} {113}},\ \bibinfo {pages}
  {247001} (\bibinfo {year} {2014})}\BibitemShut {NoStop}%
\bibitem [{\citenamefont {Hays}\ \emph {et~al.}(2018)\citenamefont {Hays},
  \citenamefont {de~Lange}, \citenamefont {Serniak}, \citenamefont {van
  Woerkom}, \citenamefont {Bouman}, \citenamefont {Krogstrup}, \citenamefont
  {Nyg\aa{}rd}, \citenamefont {Geresdi},\ and\ \citenamefont
  {Devoret}}]{PhysRevLett.121.047001}%
  \BibitemOpen
  \bibfield  {author} {\bibinfo {author} {\bibfnamefont {M.}~\bibnamefont
  {Hays}}, \bibinfo {author} {\bibfnamefont {G.}~\bibnamefont {de~Lange}},
  \bibinfo {author} {\bibfnamefont {K.}~\bibnamefont {Serniak}}, \bibinfo
  {author} {\bibfnamefont {D.~J.}\ \bibnamefont {van Woerkom}}, \bibinfo
  {author} {\bibfnamefont {D.}~\bibnamefont {Bouman}}, \bibinfo {author}
  {\bibfnamefont {P.}~\bibnamefont {Krogstrup}}, \bibinfo {author}
  {\bibfnamefont {J.}~\bibnamefont {Nyg\aa{}rd}}, \bibinfo {author}
  {\bibfnamefont {A.}~\bibnamefont {Geresdi}}, \ and\ \bibinfo {author}
  {\bibfnamefont {M.~H.}\ \bibnamefont {Devoret}},\ }\bibfield  {title}
  {\enquote {\bibinfo {title} {{Direct Microwave Measurement of
  Andreev-Bound-State Dynamics in a Semiconductor-Nanowire Josephson
  Junction}},}\ }\href {\doibase 10.1103/PhysRevLett.121.047001} {\bibfield
  {journal} {\bibinfo  {journal} {Phys. Rev. Lett.}\ }\textbf {\bibinfo
  {volume} {121}},\ \bibinfo {pages} {047001} (\bibinfo {year}
  {2018})}\BibitemShut {NoStop}%
\bibitem [{\citenamefont {{Karzig}}\ \emph {et~al.}(2017)\citenamefont
  {{Karzig}}, \citenamefont {{Knapp}}, \citenamefont {{Lutchyn}}, \citenamefont
  {{Bonderson}}, \citenamefont {{Hastings}}, \citenamefont {{Nayak}},
  \citenamefont {{Alicea}}, \citenamefont {{Flensberg}}, \citenamefont
  {{Plugge}}, \citenamefont {{Oreg}} \emph {et~al.}}]{2017PhRvB..95w5305K}%
  \BibitemOpen
  \bibfield  {author} {\bibinfo {author} {\bibfnamefont {T.}~\bibnamefont
  {{Karzig}}}, \bibinfo {author} {\bibfnamefont {C.}~\bibnamefont {{Knapp}}},
  \bibinfo {author} {\bibfnamefont {R.~M.}\ \bibnamefont {{Lutchyn}}}, \bibinfo
  {author} {\bibfnamefont {P.}~\bibnamefont {{Bonderson}}}, \bibinfo {author}
  {\bibfnamefont {M.~B.}\ \bibnamefont {{Hastings}}}, \bibinfo {author}
  {\bibfnamefont {C.}~\bibnamefont {{Nayak}}}, \bibinfo {author} {\bibfnamefont
  {J.}~\bibnamefont {{Alicea}}}, \bibinfo {author} {\bibfnamefont
  {K.}~\bibnamefont {{Flensberg}}}, \bibinfo {author} {\bibfnamefont
  {S.}~\bibnamefont {{Plugge}}}, \bibinfo {author} {\bibfnamefont
  {Y.}~\bibnamefont {{Oreg}}},  \emph {et~al.},\ }\bibfield  {title} {\enquote
  {\bibinfo {title} {{Scalable designs for quasiparticle-poisoning-protected
  topological quantum computation with Majorana zero modes}},}\ }\href
  {\doibase 10.1103/PhysRevB.95.235305} {\bibfield  {journal} {\bibinfo
  {journal} {Phys. Rev. B}\ }\textbf {\bibinfo {volume} {95}},\ \bibinfo {eid}
  {235305} (\bibinfo {year} {2017})},\ \Eprint
  {http://arxiv.org/abs/1610.05289} {arXiv:1610.05289 [cond-mat.mes-hall]}
  \BibitemShut {NoStop}%
\bibitem [{\citenamefont {Trif}\ and\ \citenamefont
  {Simon}(2019)}]{PhysRevLett.122.236803}%
  \BibitemOpen
  \bibfield  {author} {\bibinfo {author} {\bibfnamefont {Mircea}\ \bibnamefont
  {Trif}}\ and\ \bibinfo {author} {\bibfnamefont {Pascal}\ \bibnamefont
  {Simon}},\ }\bibfield  {title} {\enquote {\bibinfo {title} {{Braiding of
  Majorana Fermions in a Cavity}},}\ }\href {\doibase
  10.1103/PhysRevLett.122.236803} {\bibfield  {journal} {\bibinfo  {journal}
  {Phys. Rev. Lett.}\ }\textbf {\bibinfo {volume} {122}},\ \bibinfo {pages}
  {236803} (\bibinfo {year} {2019})}\BibitemShut {NoStop}%
\bibitem [{\citenamefont {Lubashevsky}\ \emph {et~al.}(2012)\citenamefont
  {Lubashevsky}, \citenamefont {Lahoud}, \citenamefont {Chashka}, \citenamefont
  {Podolsky},\ and\ \citenamefont {Kanigel}}]{Lubashevsky2012}%
  \BibitemOpen
  \bibfield  {author} {\bibinfo {author} {\bibfnamefont {Y.}~\bibnamefont
  {Lubashevsky}}, \bibinfo {author} {\bibfnamefont {E.}~\bibnamefont {Lahoud}},
  \bibinfo {author} {\bibfnamefont {K.}~\bibnamefont {Chashka}}, \bibinfo
  {author} {\bibfnamefont {D.}~\bibnamefont {Podolsky}}, \ and\ \bibinfo
  {author} {\bibfnamefont {A.}~\bibnamefont {Kanigel}},\ }\bibfield  {title}
  {\enquote {\bibinfo {title} {{Shallow pockets and very strong coupling
  superconductivity in FeSe$_x$Te$_{1-x}$}},}\ }\href {\doibase
  10.1038/nphys2216} {\bibfield  {journal} {\bibinfo  {journal} {Nature
  Physics}\ }\textbf {\bibinfo {volume} {8}},\ \bibinfo {pages} {309--312}
  (\bibinfo {year} {2012})}\BibitemShut {NoStop}%
\bibitem [{\citenamefont {Rinott}\ \emph {et~al.}(2017)\citenamefont {Rinott},
  \citenamefont {Chashka}, \citenamefont {Ribak}, \citenamefont {Rienks},
  \citenamefont {Taleb-Ibrahimi}, \citenamefont {Fevre}, \citenamefont
  {Bertran}, \citenamefont {Randeria},\ and\ \citenamefont
  {Kanigel}}]{doi:10.1126/sciadv.1602372}%
  \BibitemOpen
  \bibfield  {author} {\bibinfo {author} {\bibfnamefont {Shahar}\ \bibnamefont
  {Rinott}}, \bibinfo {author} {\bibfnamefont {K.~B.}\ \bibnamefont {Chashka}},
  \bibinfo {author} {\bibfnamefont {Amit}\ \bibnamefont {Ribak}}, \bibinfo
  {author} {\bibfnamefont {Emile D.~L.}\ \bibnamefont {Rienks}}, \bibinfo
  {author} {\bibfnamefont {Amina}\ \bibnamefont {Taleb-Ibrahimi}}, \bibinfo
  {author} {\bibfnamefont {Patrick~Le}\ \bibnamefont {Fevre}}, \bibinfo
  {author} {\bibfnamefont {François}\ \bibnamefont {Bertran}}, \bibinfo
  {author} {\bibfnamefont {Mohit}\ \bibnamefont {Randeria}}, \ and\ \bibinfo
  {author} {\bibfnamefont {Amit}\ \bibnamefont {Kanigel}},\ }\bibfield  {title}
  {\enquote {\bibinfo {title} {{Tuning across the BCS-BEC crossover in the
  multiband superconductor Fe$_{1+y}$Se$_x$Te$_{1-x}$: An angle-resolved
  photoemission study}},}\ }\href {\doibase 10.1126/sciadv.1602372} {\bibfield
  {journal} {\bibinfo  {journal} {Science Advances}\ }\textbf {\bibinfo
  {volume} {3}},\ \bibinfo {pages} {e1602372} (\bibinfo {year} {2017})},\
  \Eprint
  {http://arxiv.org/abs/https://www.science.org/doi/pdf/10.1126/sciadv.1602372}
  {https://www.science.org/doi/pdf/10.1126/sciadv.1602372} \BibitemShut
  {NoStop}%
\bibitem [{\citenamefont {Xu}\ \emph {et~al.}(2023)\citenamefont {Xu},
  \citenamefont {Wong}, \citenamefont {Mascot},\ and\ \citenamefont
  {Morr}}]{PhysRevB.107.214514}%
  \BibitemOpen
  \bibfield  {author} {\bibinfo {author} {\bibfnamefont {Chang}\ \bibnamefont
  {Xu}}, \bibinfo {author} {\bibfnamefont {Ka~Ho}\ \bibnamefont {Wong}},
  \bibinfo {author} {\bibfnamefont {Eric}\ \bibnamefont {Mascot}}, \ and\
  \bibinfo {author} {\bibfnamefont {Dirk~K.}\ \bibnamefont {Morr}},\ }\bibfield
   {title} {\enquote {\bibinfo {title} {{Competing topological superconducting
  phases in ${\mathrm{FeSe}}_{0.45}{\mathrm{Te}}_{0.55}$}},}\ }\href {\doibase
  10.1103/PhysRevB.107.214514} {\bibfield  {journal} {\bibinfo  {journal}
  {Phys. Rev. B}\ }\textbf {\bibinfo {volume} {107}},\ \bibinfo {pages}
  {214514} (\bibinfo {year} {2023})}\BibitemShut {NoStop}%
\end{thebibliography}%

\clearpage
\newpage

\setcounter{equation}{0}
\setcounter{figure}{0}
\setcounter{section}{0}
\setcounter{table}{0}
\setcounter{page}{1}
\makeatletter
\renewcommand{\theequation}{S\arabic{equation}}
\renewcommand{\thepage}{S-\arabic{page}}
\renewcommand{\thesection}{S\arabic{section}}
\renewcommand{\thefigure}{S\arabic{figure}}
\renewcommand{\thetable}{S-\Roman{table}}

\begin{widetext}
\begin{center}
Supplementary materials on \\
\textbf{``Microwave spectroscopy of Majorana vortex modes''}\\
Zhibo Ren$^{1}$, Justin Copenhaver$^{1,2}$, Leonid Rokhinson$^{1}$, Jukka I. V{\"a}yrynen$^{1}$, \\ 
$^{1}$ \textit{Department of Physics and Astronomy, Purdue University, West Lafayette, Indiana 47907, USA}\\
$^{2}$ \textit{Department of Physics, University of Colorado, Boulder, CO 80309, USA}\\
\end{center}

These supplementary materials contain details about the critical cavity Q-factor and screened potential of a vortex.

\section{Derivation of the critical cavity Q-factor }
In the strong coupling limit, where the MW coupling strength greatly exceeds the vortex bound state level width $\delta$, we can approximate the correlation function in Eq.~(\ref{eq:correlation function}) by neglecting its imaginary part. The approximation leads to an expression for the correlation function given by (take $l=1$)
\begin{equation}
\Pi^{(p)}(\omega) \approx \frac{2\omega_{1}^{(p)}}{(\omega_{1}^{(p)})^2-\omega^{2}}[|q_{1,\text{M}}|^2 n_\text{M}+|q_{1,-\text{M}}|^2(1-n_\text{M})],
\label{eq: real correlation function}
\end{equation}
where $n_\text{M}$ is the Majorana occupation number and $p = (-1)^{n_\text{M}}$ its parity. 
The resonator transmission,  Eq.~(\ref{eq:transmission}) of the main text, then takes the form 
\begin{equation}
    |\tau^{(p)}(\omega)|^2 \approx\frac{\kappa^2}{[\omega-\omega_c+\frac{ \omega_c \Pi^{(p)}(\omega)}{2 C_{\text{tot}}}]^2+\kappa^2}.
    \label{eq: abs transmission}
\end{equation}
We can read off the modified resonance frequencies from the 
equation, 
\begin{equation}
\omega-\omega_{c}+\frac{\omega_c \Pi^{(p)}(\omega)}{2 C_{\text{tot}}}=0.
\label{eq: peak equation}
\end{equation}
Next, we solve this equation in both the resonant limit and the detuning limit, and find that there are always 3 solutions, we denote them as $\Omega_{n}^{(p)},\Omega_{c}^{(p)}$, and $\Omega_{t}^{(p)}$.

In the resonant region $|\omega_c-E_1|\ll \omega_c \zeta$, these solutions are given by 
\begin{equation}
\begin{cases}
\Omega_{n}^{(p)} & \approx-\omega_{1}^{(p)}+\frac{(\omega_{c}-\omega_{1}^{(p)})}{2}[\zeta_{+\text{M}} n_\text{M}+\zeta_{-\text{M}}(1-n_\text{M})],\\
\Omega_{c}^{(p)} & \approx\frac{1}{2}(\omega_{c}+\omega_{1}^{(p)})-\omega_c[\zeta_{+\text{M}} n_\text{M}+\zeta_{-\text{M}}(1-n_\text{M})],\\
\Omega_{t}^{(p)} & \approx\frac{1}{2}(\omega_{c}+\omega_{1}^{(p)})+\omega_c[\zeta_{+\text{M}} n_\text{M}+\zeta_{-\text{M}}(1-n_\text{M})].
\label{eq: peak solution1}
\end{cases}
\end{equation}
The first solution $\Omega_{n}^{(p)}$ is negative, so the peak with the lowest positive frequency of the transmission is located at $\Omega_{c}^{(p)}$, with a peak width approximately equal to $\kappa$. The shift of the peak positions, denoted as $\Delta \Omega_c$, is given by
\begin{equation}
\Delta \Omega_c=\Omega_{c}^{(+)}-\Omega_{c}^{(-)} \approx E_{\text{M}}+\omega_c \delta \zeta.
\label{eq: peak shift1}
\end{equation}
Approximate to the zero-order of $\zeta$, the average of peak positions $\Omega_{c}=\frac{1}{2}(\Omega_{c}^{(+)}+\Omega_{c}^{(-)})\approx \omega_c$.
 The critical cavity Q-factor is achieved when the shift in peak position equals the escape rate, $\Delta \Omega_c=\kappa$. This lead to the expression for $Q_{c}$,
\begin{equation}
Q_{c}=\frac{\omega_c}{|E_{\text{M}}+\omega_c \delta \zeta|}, 
\label{eq: Qc1}
\end{equation}
giving Eq.~(\ref{eq:Qc1}) of the main text.

In the detuning region $|E_1-\omega_c| \gg \omega_c\zeta, E_{\text{M}}$, the Eq.~(\ref{eq: peak equation}) has 3 solutions given by
\begin{equation}
\begin{cases}
\Omega_{n}^{(p)} & \approx -\omega_{1}^{(p)}-\frac{\omega_c^2}{(\omega_{c}+\omega_{1}^{(p)})}[\zeta^2_{+\text{M}} n_\text{M}+\zeta^2_{-\text{M}}(1-n_\text{M})],\\
\Omega_{c}^{(p)} & \approx \omega_{c}-\frac{2\omega_{1}^{(p)}\omega_c^2}{|\omega_{c}^2-(\omega_{1}^{(p)})^2|}[\zeta^2_{+\text{M}} n_\text{M}+\zeta^2_{-\text{M}}(1-n_\text{M})],\\
\Omega_{t}^{(p)} & \approx \omega_{1}^{(p)}-\frac{\omega_c^2}{|\omega_{c}-\omega_{1}^{(p)}|}[\zeta^2_{+\text{M}} n_\text{M}+\zeta^2_{-\text{M}}(1-n_\text{M})].
\label{eq: peak solution2}
\end{cases}
\end{equation}
Similar to the resonant region, the peak with the lowest positive frequency of the transmission is located at $\Omega_{c}^{(p)}$, and the peak width is now doubled to $2\kappa$. The shift of the peak positions is given by
\begin{equation}
\Delta \Omega_c=\Omega_{c}^{(+)}-\Omega_{c}^{(-)} \approx \frac{4(E_1^2+\omega_c^2)\omega_c^2}{(E_1^2-w_c^2)^2} E_{\text{M}}\zeta^2-\frac{4E_1\omega_c^2}{|E_1^2-w_c^2|}\zeta \delta \zeta.
\label{eq: peak shift2}
\end{equation}
Approximate to the zero-order of $\zeta$, the average of peak positions $\Omega_{c}=\frac{1}{2}(\Omega_{c}^{(+)}+\Omega_{c}^{(-)})\approx \omega_c$
 The critical cavity Q-factor is achieved when the shift in peak position equals the escape rate, $\Delta \Omega_c=\kappa$.
\begin{equation}
Q_{c}=\frac{(E_1^2-w_c^2)^2}{4\omega_c\zeta|(E_1^2+\omega_c^2) E_{\text{M}}\zeta-E_1(E_1^2-w_c^2)\delta \zeta|},
\label{eq: Qc2}
\end{equation}
leading to Eq.~(\ref{eq:Qc2a},\ref{eq:Qc2b}) of the main text.

\section{Screening of a vortex }
Here we model the screened potential of a vortex based on 3D parabolic bulk bands~\cite{PhysRevLett.77.566,Lubashevsky2012,doi:10.1126/sciadv.1602372,PhysRevB.107.214514}. The opening of a gap $\Delta(R)$ in the spectrum results in a displacement of the carrier density by $\delta n(R)$, corresponding to a (non-screened) charge density $\rho(R) = -e\,\delta n(R)$. For a single vortex of size $\xi$ with order parameter given by Eq.~(\ref{eq: single vortex}), 
\begin{equation}
    \rho(R) = eN_\mu\Delta_0^2 \frac{\xi^2}{R^2 + \xi^2} \frac{d\,\text{ln}\,T_c}{d\mu} \,,
    \label{Charge density profile}
\end{equation}
with $R$ the radial distance from 
the vortex core and $N_\mu\,d\,\text{ln}\,T_c/d\mu \approx n/\mu^2$~\cite{PhysRevLett.77.566}, where $n$ is the electron density and $N_\mu$ is the density of states at chemical potential $\mu$. With two vortices separated by a distance $d > \xi$, the charge density contains contributions from each vortex.

To account for electric screening due to bulk carriers~\cite{Lubashevsky2012,doi:10.1126/sciadv.1602372} in the  superconductor, we use a Thomas-Fermi approximation and solve the screened Poisson equation for the screened electric potential $\varphi_{\text{scrn}}(R)$ ~\cite{PhysRevLett.77.566}:
\begin{equation}
    [\nabla^2 - \lambda_{\text{TF}}^{-2}]\,\varphi_{\text{scrn}} = -4\pi\rho \,,
    \label{eq:Screened Poisson equation}
\end{equation}
where $\lambda_{\text{TF}}=(8\pi e^2 N_\mu)^{-1/2}$ 
is the screening length. In our numerical simulation, we take $\lambda_{\text{TF}}$ to be one lattice constant, $\lambda_{\text{TF}}\approx 5$ nm. 
The Green's function corresponding to Eq.~(\ref{eq:Screened Poisson equation}) is $G(\Vec{r},\Vec{r}\,') = -\frac{1}{4\pi\abs{\Vec{r}-\Vec{r}\,'}} e^{-\abs{\Vec{r}-\Vec{r}\,'}/\lambda_{\text{TF}}}$, therefore, on the lattice we have
\begin{equation}
    \varphi_{\text{scrn}}(\Vec{r}) = \int_V \frac{\rho(\Vec{r}\,')}{\abs{\Vec{r} - \Vec{r}\,'}} e^{-\abs{\Vec{r}-\Vec{r}\,'}/\lambda_{\text{TF}}}d^3 r\,'.
    \label{eq:Screened electric potential}
\end{equation}
After performing this calculation with a two-vortex source term, we then use $\varphi_{\text{scrn}}$ in Hamiltonian given by Eq.~(\ref{eq:3D SC}), by setting $\mu \to \mu + e\varphi_{\text{scrn}}$. 

\begin{figure}
\centering
\includegraphics[width=1\columnwidth]{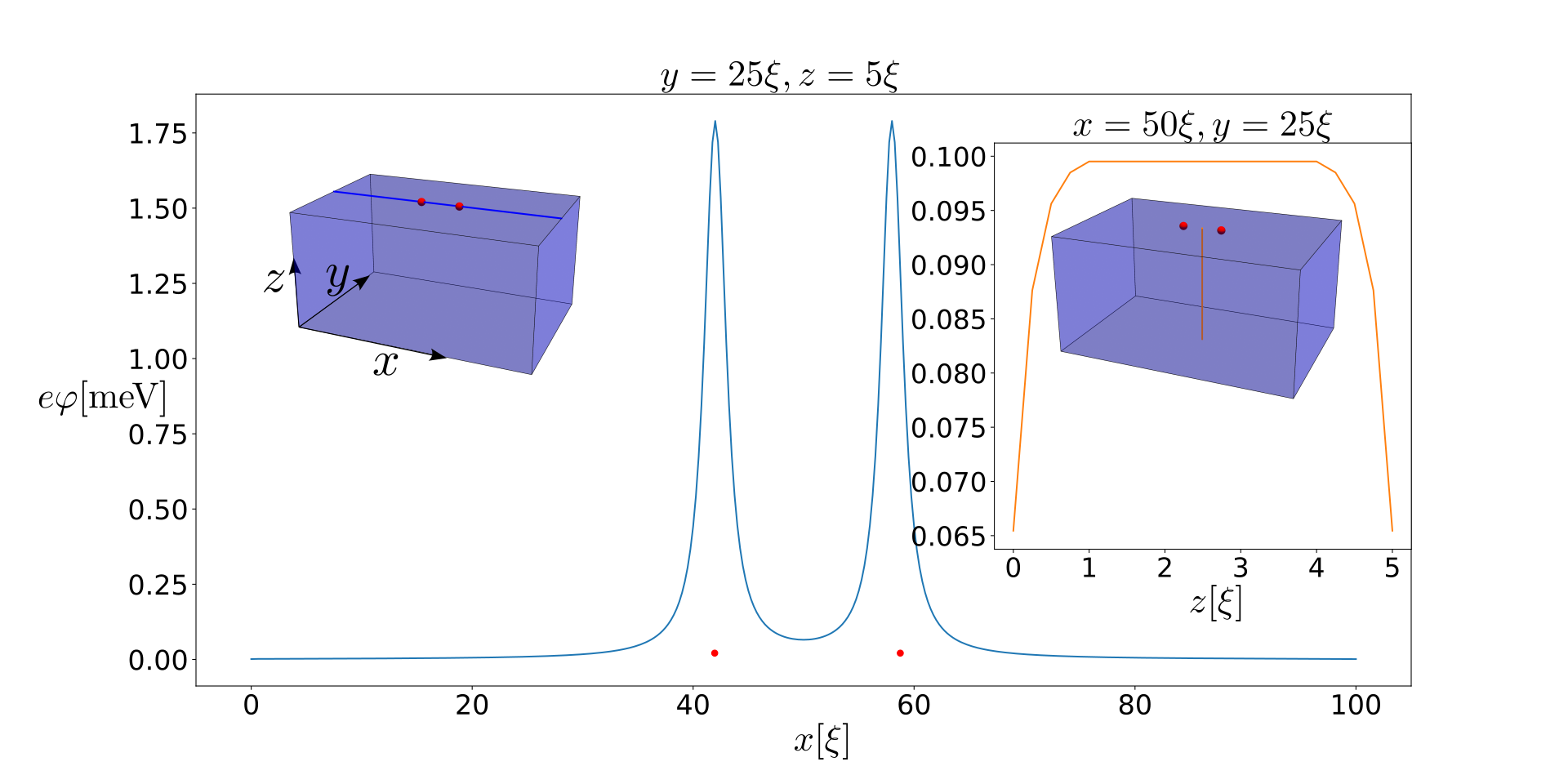}

\caption{
The screened potential in x and z (inset) directions for a system with two vortices and $\lambda_{\text{TF}}=a =\xi$. The system is a cuboid with dimensions $100\xi \cross 50\xi \cross 5\xi$. 
The two red dots indicate the positions of the two vortices. 
The screened potential in the x direction is plotted along the blue line in the left cuboid and in the z direction (inset) it is plotted along the orange line in the right cuboid. 
} \label{fig:screened potential}
\end{figure}

\end{widetext}

\end{document}